\documentclass[amsmath,amssymb,twocolumn,aps]{revtex4-1}

\usepackage{graphicx}
\usepackage{dcolumn}
\usepackage{bm}
\usepackage{floatrow}
\usepackage{comment}
\usepackage[caption=false]{subfig}

\usepackage[table,xcdraw]{xcolor}
\usepackage{amsmath}
\usepackage{graphicx}
\usepackage[colorinlistoftodos]{todonotes}
\usepackage[colorlinks=true, allcolors=blue]{hyperref}

\usepackage{bbm}
\usepackage{amsfonts}
\usepackage{multirow}
\usepackage{amssymb}
\usepackage{natbib}
\usepackage{enumitem}

\newcommand{\beginsupplement}{%
        \setcounter{table}{0}
        \renewcommand{\thetable}{S\arabic{table}}%
        \setcounter{figure}{0}
        \renewcommand{\thefigure}{S\arabic{figure}}%
     }

\begin{document}
\captionsetup[subfigure]{labelformat=empty}

\preprint{APS/123-QED}

\title{Large algebraic connectivity fluctuations in spatial network ensembles imply a predictive advantage from node location information}

\author{Matthew Garrod}
\author{Nick S. Jones}%

\affiliation{%
Department of Mathematics, Imperial College London.
}%

\date{\today}

\begin{abstract}
A Random Geometric Graph (RGG) ensemble is defined by the disordered distribution of its node locations. We investigate how this randomness drives sample-to-sample fluctuations in the dynamical properties of these graphs. We study the distributional properties of the algebraic connectivity which is informative of diffusion and synchronization timescales in graphs. We use numerical simulations to provide the first characterisation of the algebraic connectivity distribution for RGG ensembles. We find that the algebraic connectivity can show fluctuations relative to its mean on the order of $30 \%$, even for relatively large RGG ensembles ($N=10^5$). We explore the factors driving these fluctuations for RGG ensembles with different choices of dimensionality, boundary conditions and node distributions. Within a given ensemble, the algebraic connectivity can covary with the minimum degree and can also be affected by the presence of density inhomogeneities in the nodal distribution. We also derive a closed-form expression for the expected algebraic connectivity for RGGs with periodic boundary conditions for general dimension. 
\end{abstract}

\pacs{Valid PACS appear here}
\maketitle
\section{\label{sec:level1}Introduction}

\subsection{Motivation} \label{Backgroun}

Networks frequently show spatial or metric structure where nodes possessing similar attributes are more likely to share a connection \cite{barthelemy_spatial_2011,barnett_spatially_2007}. A general framework for modelling networks where edge formation is dependent on continuous node attributes is the Soft Random Geometric Graph \cite{penrose2016connectivity,dettmann_random_2016,giles2016connectivity} or Spatially Embedded Random Network \cite{barnett_spatially_2007}. In these models node positions, $X_i$, for $i=1,...,N$ are sampled in some chosen domain, $\mathcal{D}$, according to some distribution $P(\cdot)$. Connections between nodes are then drawn independently with a probability which is a function of their Euclidean distance. This is typically expressed via some connection probability function $f(r)$ which specifies the probability that two nodes separated by a distance, $r$, are connected. 

Models of within this class have been used in a diverse range of applications including: social networks \cite{butts_spatial_2011,butts_interorganizational_2012,daraganova_networks_2012}, wireless communications networks \cite{dettmann_random_2016,glauche_continuum_2003,kenniche_random_2010}, spatially constrained networks in the brain \cite{odea_spreading_2013,lo_geometric_2015} and transport networks \cite{hackl2017generation}. However, in practice, the amount of information that we possess about the graph structure can vary significantly. For example, in some cases, we might posses microscopic data about individual nodes, such as, geographic locations \cite{daraganova_networks_2012,butts_interorganizational_2012} or socio-economic coordinates \cite{leo2016socioeconomic} of people or organisations in a social network. In other cases we may possess coarse-grained data describing the density of individuals. For instance, population density contained within census data which has been used as an input for models of social networks \cite{hipp2013extrapolative,butts_geographical_2012} and transportation networks \cite{hackl2017generation}. Furthermore, even when we do not possess data about individual nodes, we can often obtain their relative positions in some latent embedding space \cite{hoff2002latent}. In fact, it has been found that many complex networks can be well represented by embedding the nodes in a hyperbolic space \cite{krioukov_hyperbolic_2010,papadopoulos2012popularity,kleineberg2016hidden}.

We can expect higher levels of detail to be more costly:  in social networks, conducting a survey to obtain microscopic details about individuals will carry some cost, whereas census data containing distributions of individuals traits across populations may be freely available. In a particular situation it becomes relevant to ask: \emph{what level of information do we need to know about the node attributes in order to determine the structural or dynamical properties of the network to the desired degree of accuracy?} In this paper we answer the above question by studying ensemble variability in spatial network ensembles. By ensemble variability we refer to the degree to which different graphs drawn from the same random graph ensemble vary in their structural and dynamical properties.

\begin{figure}[h]
\centering
\includegraphics[width=0.95\textwidth]{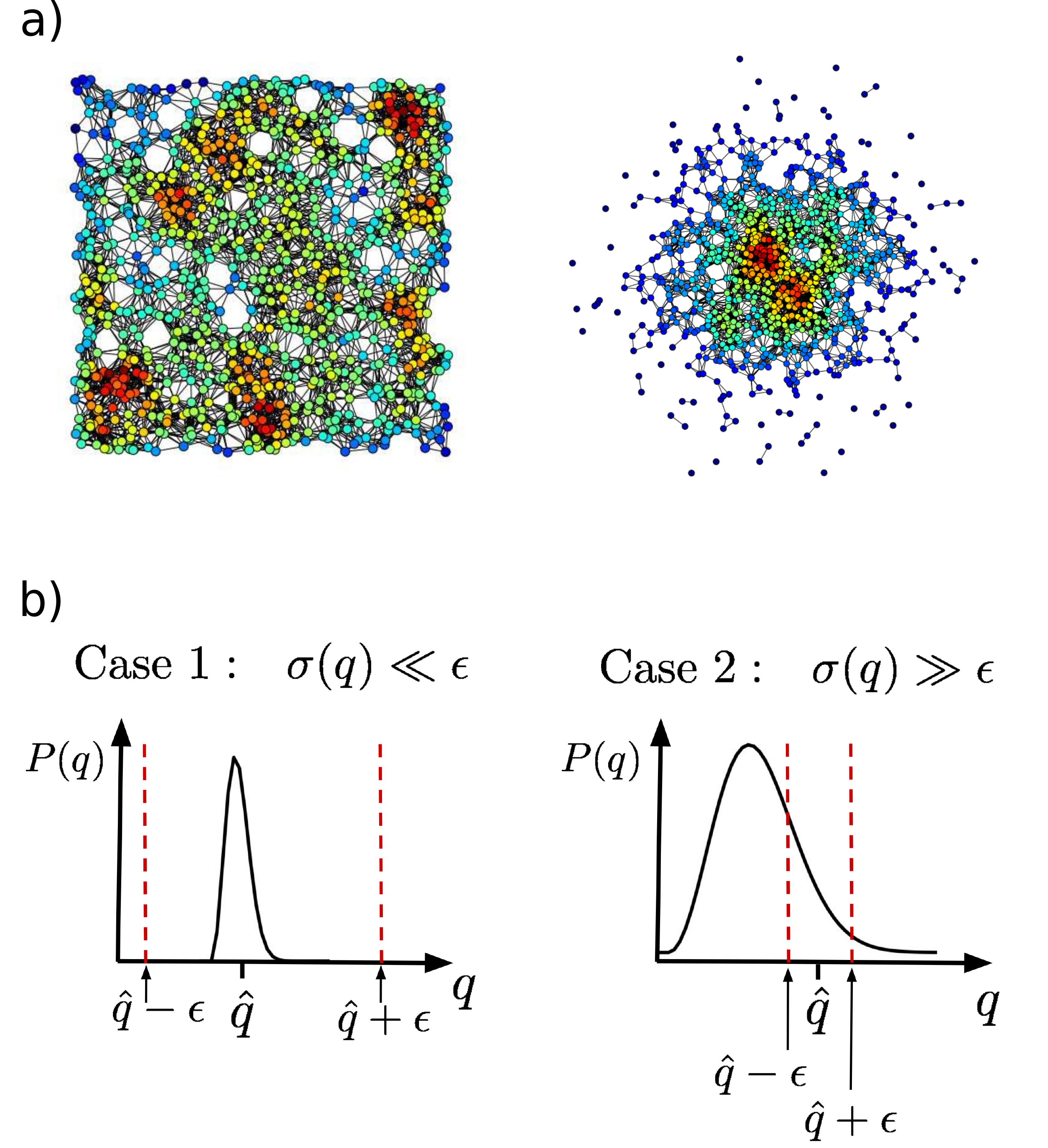}
\caption{a) Examples of RGGs with $N=1000$ for different node distributions. Shows the case of a uniform node distribution in $[0,1]^2$ with $R=0.082$ (left) and  2D Gaussian node distribution in $\mathbb{R}^2$ with unit variance and zero mean with $R=0.281$ (right). b) Figure illustrating how ensemble variability of a graph property, $q$, relates to our ability to be able to estimate it to within some threshold value, $\epsilon$. Case 1 shows an example where the ensemble variability of $q$ is much smaller than the desired precision, $\epsilon$. Case 2 illustrates the case where the ensemble variability of $q$ is much larger than the desired precision. In the former case it will be possible to make estimates of the true value of $q$ to the desired precision by taking the value of $q$ from any graph sampled from the ensemble.}
\label{RGG_Example}
\end{figure}

The choice of $f(r)$ varies widely between different applications \cite{butts_interorganizational_2012,daraganova_networks_2012,lo_geometric_2015,odea_spreading_2013}. We focus on the case of the Random Geometric Graph (RGG) for which two nodes $i$ and $j$ are connected iff $|X_i-X_j|\leq R$ (Figure \ref{RGG_Example}). Studying ensemble variability in RGGs allows us to characterise how randomness in the node locations influences variability in the structural and dynamical properties of the network. This in turn allows us to identify the level of knowledge (e.g node distribution or precise knowledge of node locations) required to make accurate predictions of structurally and dynamically relevant properties.

\subsection{Problem Setup}

Suppose we are interested in a network property, $q$. For a particular random geometric graph ensemble $q$ will be a random variable with an expected value $\mathbb{E}(q)$, standard deviation $\sigma(q)$ and distribution $\mathcal{P}_{\textit{q}}(q)$. Given a distribution, $P(\cdot)$, $N$ and $R$ it is possible to draw representative members from the ensemble in order to estimate $q$. For some RGG ensembles, we may also be able to estimate $\mathbb{E}(q)$ analytically given knowledge of the node distribution and connection radius. 

Suppose the true value of $q$ for a graph is given by $q^*$. In practice we may want to predict the value of $q$ for a particular RGG to within some threshold $\epsilon$ ie. the predicted value $\tilde{q}$ satisfies: $\tilde{q} \in [ q^* + \epsilon , q^* - \epsilon]$. Whether this is possible given knowledge of $P(\cdot)$ alone will depend on the standard deviation of the distribution of $q$. For example, there are two limiting cases (See Figure \ref{RGG_Example}b):
\begin{enumerate}[leftmargin=*]
\item{\textbf{Case 1:} If $\sigma(q) << \epsilon$, the vast majority of values of $q$ sampled from $\mathcal{P}_{\textit{q}}(q)$ will lie within the desired region. As a consequence, a sensible guess for $q^*$ could be made by taking $\mathbb{E}(q)$.} 
\item{\textbf{Case 2:} If $\sigma(q) >> \epsilon$, then samples from the distribution $\mathcal{P}_{\textit{q}}(q)$ are likely to be uninformative of the value of $q^*$ for the particular graph.}
\end{enumerate}
In the latter case knowledge of the set of node positions $\underline{X}=(X_1,X_2,...,X_N)$ will be of utility since it will allow us to generate a network for which we can compute $q$.

The particular choice of $\epsilon$ is subjective and depends on the application. In this work we investigate the case where $\epsilon \approx \mathbb{E}(q)$, that is, we are interested in when it will be possible to predict $q$ to within a threshold determined by the length scale of the mean of the distribution. 

Given the above choice of $\epsilon$, a relevant metric to study when comparing the typical width of $\mathcal{P}_{\textit{q}}(q)$ for different network ensembles is the \emph{coefficient of variation} (CV). The CV is defined as the standard deviation of a distribution divided by its mean: $CV(q) = \frac{\sigma(q)}{\mathbb{E}(q)}$. If $CV(q) \approx 0 $ then $\mathbb{E}(q)$ will be sufficient to predict $q$ for most ensemble members. In contrast, if $CV(q)$ is relatively large, then knowledge of $\mathbb{E}(q)$ will not be sufficient in order to make precise predictions about the value of $q$ for a particular ensemble member. 

The notion of whether a network property is well-represented by its mean across random network ensembles has been studied in the past in \cite{kim_ensemble_2007} and \cite{carlson_sample--sample_2011}. They refer to a property which can be well represented by its ensemble average as being \emph{ensemble averageable}. They studied the degree to which certain spectral properties of networks are ensemble averageable in scale-free and real world network ensembles. A property is strictly ensemble averageable if $\sigma(q)$ remains finite in the thermodynamic limit. However, since most real world networks are large but finite we instead focus on identifying cases where $CV(q)$ remains large (for example fluctuations on the order $5-10 \%$ at least) for the ranges of $N$ studied.

\subsection{The Algebraic Connectivity}

Networks can be studied via the adjacency matrix, $A$, for which $A_{ij} = 1$ if nodes $i$ and $j$ share an edge and $0$ otherwise, and the Laplacian matrix, $L$, which has elements $L_{ij} = \delta_{ij} \big(\sum_j A_{ij} \big) - A_{ij}$. The smallest non-zero eigenvalue of this matrix, $\mu_2$, is commonly known as the \emph{algebraic connectivity}. This quantity is connected to both dynamical properties such as the characteristic timescale of diffusion \cite{newman_networks:_2010} and synchronizability \cite{estrada_synchronizability_2015,dorfler_synchronization_2013} as well as a metric of how difficult it is to partition the network into two components \cite{donetti_optimal_2006}. In the study of wireless communications networks, the algebraic connectivity is related to the time required for linear consensus algorithms to run to completion and their total energy consumption \cite{barbarossa_achieving_2007,sardellitti_optimal_2012,olfati-saber_ultrafast_2005}. As a consequence, there are numerous cases in which we might want to estimate the algebraic connectivity of large networks. 

The algebraic connectivity of RGGs has been studied in \cite{estrada_consensus_2016} where they find a bound on its value in terms of the number of nodes and the domain size. In addition, in \cite{sardellitti_optimal_2012} they derive an analytic approximation for the algebraic connectivity of $2d$ RGGs with periodic boundary conditions. However, to our knowledge, there have been no studies of the distribution of $\mu_2$ values, $\mathbb{P}(\mu_2)$, in RGG ensembles with fixed parameter values. We will also consider the effect of varying the spatial dimension of the system as many complex networks, such as large scale social networks, can be represented using a moderate number of dimensions \cite{bonato_dimensionality_2014}.

The algebraic connectivity distribution has been studied in \cite{carlson_sample--sample_2011} for network ensembles modeled on real world networks and in \cite{peixoto_eigenvalue_2013} for stochastic block models. In both cases it is reported that the distribution of values can be `broad' or of high variance. However, neither study addresses the question of how the width and form of $\mathbb{P}(\mu_2)$ depends on the parameters of the random graph model.

The eigenvalue spectra of the adjacency matrix for RGGs has been studied in \cite{dettmann_spectral_2017,alonso_weighted_2017,dettmann_symmetric_2017,dettmann_symmetric_2017}. In the former case they show that the spectrum shows some universal features also seen in the spectral distributions of other random network models, while in the latter it is shown how the spectral properties can be related to the frequency of occurrence of certain subgraphs (also known as \emph{motifs}) in the network. However, no research so far has focused on the distributions of individual eigenvalues in RGGs which is the subject of the present manuscript.

 The algebraic connectivity of a disconnected graph is equal to zero. In this work we will consider the value of $\mu_2$ for the largest connected component (LCC) of the graph. The values of other graph properties will also correspond to those for the LCC unless otherwise specified.

In some of the following sections we study the covariation between $\mu_2$ and the other network properties. This can be quantified by computing Spearman's rank correlation. This metric measures the extent of the monotonic correlation between two variables. We will denote the Spearman correlation between two variables $X$ and $Y$ as $\rho(X,Y)$. 

\subsection{Summary of Results}

The main aim of this study was to identify regions in parameter space where $CV(\mu_2)$ is large. These regions correspond to RGG ensembles where additional information about node locations may give us a meaningful predictive advantage. Our main conclusions are:
\begin{itemize}
\item{For RGGs with homogeneous node distributions the value of $CV(\mu_2)$ is relatively small in low dimensional systems. However, for ensembles with larger values of $d$ we can observe large fluctuations in the value of $\mu_2$ which persist for relatively large graphs. For instance we observe values of $CV(\mu_2)$ greater than 0.2 in RGGs with toroidal boundaries with $d>8$ and RGGs with solid boundaries with $d>5$ (Section \ref{dimvar}).}
\item{The main factor governing the behaviour of $\mu_2$ in higher dimensional RGGs with homogeneous node distributions is the minimum degree, $\kappa_{\mathrm{min}}$ (Sections \ref{av_mu2} and \ref{dimvar}). This conclusion holds so long as $R$ is sufficient for the RGGs to have $\kappa_{\mathrm{min}} \geq 1$.}
\item{The presence of low density regions in the node distribution can also lead to significant fluctuations in $\mu_2$. For instance, 2d RGG ensembles with Gaussian node distributions with $N=10^5$ have values of $CV(\mu_2)$ greater than 0.25 for a range of parameters (Section \ref{nonuniform}). These fluctuations are much more significant than those in homogenous RGGs with comparable average degrees. In this case the fluctuations are driven by the presence of weakly connected subgrahs and bottlenecks which occur as chance events in different graphs drawn from the same ensemble.}
\end{itemize}
We also demonstrate how the results can be interpreted in terms of the localisation properties of the eigenvector associated with $\mu_2$ (Section \ref{Fied_Sect}).

\section{Methodology} \label{Methods}

\subsection{Generation of RGG Ensembles with Given Expected Degree}

In order to compare different RGG ensembles it is helpful to be able to specify the mean degree of the ensemble. In this section we assume that node positions lie in the domain $[0,1]^d$ with either solid or periodic boundaries. In the high density limit the mean degree of the network can be computed by estimating the number of points which fall within the connection radius of a randomly chosen node. For the case where the domain of interest is of unit volume the probability of some node $j$ falling within the connection radius of a node $i$ is given by the volume of the ball of radius $R$ in $d$ dimensions. Multiplying this quantity by the number of remaining nodes in the network gives us an estimate for the mean degree of the form:
\begin{equation} \label{MeanDeg}
\tilde{\kappa} = (N-1) \frac{ \pi^{\frac{d}{2}}}{ \Gamma( \frac{d+2}{2} ) } R^d .
\end{equation}
Inverting this formula allows us to obtain an approximate expression for the connection radius required to generate RGG ensembles with true mean degree, $\kappa$, in $d$ dimensions for a network with $N$ nodes:
\begin{equation} \label{RadEq}
R = \frac{1}{\sqrt[]{\pi}} \bigg( \frac{\kappa}{N-1} \Gamma\bigg( \frac{d+2}{2} \bigg)   \bigg)^{\frac{1}{d}} .
\end{equation}
This approach works well in practice for RGGs with periodic boundary conditions. However, for RGGs generated in $[0,1]^d$ with solid boundaries the true mean degree of the ensemble, $\kappa$, will in general be lower than the value of $\tilde{\kappa}$ due to the presence of isolated nodes at the boundaries. In theory it is possible to compute the radius required to obtain a given expected mean degree using techniques described in \cite{dettmann_random_2016} and \cite{coon_full_2012}. However, this problem becomes analytically intractable for higher dimensional systems as the number of boundary, edge and corner terms in the integration will increase dramatically. Consequently, we must rely on a different procedure to generate RGGs with the desired mean degree for the solid boundary case. 

Equation \eqref{RadEq} can also perform poorly for higher dimensional systems with periodic boundaries. If $R>0.5$ for a node embedded in $[0,1]^d$ then the connectivity radius will cross over the periodic boundary and overlap with itself. As a consequence, the volume contained within the `ball of connectivity' will be smaller than that required to obtain the desired mean degree. For small $d$ this is not usually an issue since we require a large value of $\kappa$ for the wrap-around effect to occur. However, for the case of higher dimensions, this wrap-around effect can occur for RGGs with relatively small mean degrees. For example, if we take $N = 10^4, \kappa = 20.0 ,d = 15$ then using \eqref{RadEq} gives $R \approx 15.8$ which indicates that the `ball of connectivity' can wrap-around the periodic domain multiple times despite only filling  only a very small fraction of the entire volume. 

Given the above it is necessary to estimate the value of $R$ required to construct RGG ensembles with a given expected degree on a case by case basis. This can be achieved in practice by taking samples from the distribution of distances for each a particular choice of node distribution (see section \ref{meandeg_alg} of the SI). For each set of numerical simulations we have produced a table of radii values required to construct the RGG ensembles with mean degree values close to the required value (see section \ref{Tables} of the SI). The values of $R$ obtained in each case were validated by computing the average mean degree of the simulated RGG ensembles.

\subsection{Percolation and Connectivity}

Consider an RGG in some domain $\mathcal{D}$ with fixed, $d$, $N$ and some choice of $P(\cdot)$. As we increase $\kappa$ we will observe two significant transitions. For small $\kappa$, RGGs sampled from the ensemble will consist of small disconnected clusters of nodes. Once we pass a certain threshold we will find that a macroscopic fraction of the nodes lie within the largest connected component (LCC) of the graph. The transition to this regime is commonly referred to as the percolation transition. If we increase $\kappa$ further we will eventually reach a point where all nodes lie within the LCC with high probability. We will refer to such a graph as being connected. The dominant contribution to the connectivity probability in RGGs comes from single isolated nodes \cite{coon_full_2012}. As a consequence, the value of $\kappa$ required to achieve connectivity is highly sensitive to the shape of the domain and presence of boundaries.

In order to achieve connectivity in RGGs the value of $\kappa$ must be scaled logarithmically with $N$ \cite{penrose2003random}. This can be achieved by setting:
\begin{equation}
\kappa = C \log(N) ,
\end{equation}
where $C$ is a positive constant. The connectivity threshold in the large $N$ limit is $C=1$. Therefore, graphs with $C>1$ are highly likely to be connected, while those with $C<1$ are most likely to be disconnected. Tuning $C$ allows us to specify how far above the threshold of connectivity a given graph ensemble is.

The majority of ensembles studied in this paper are such that the number of nodes in the LCC, $N_{LCC}$, is equal to or close to $N$. We take this as justification for reporting the parameter $N$ rather than $N_{LCC}$.

\section{Ensemble Variability in the Algebraic Connectivity} \label{EnsAverage}

In this section we explore how $\mathbb{E}(\mu_2)$ and $CV(\mu_2)$ vary as a function of the system parameters for different RGG ensembles. We consider three distinct RGG ensembles:
\begin{enumerate}[leftmargin=*]
\item{\textbf{Periodic Boundaries:} Node positions drawn from a uniform distribution in $[0,1]^d$ with toroidal boundary conditions. }
\item{\textbf{Solid Boundaries:} Node positions drawn from a uniform distribution in $[0,1]^d$ with solid boundaries.}
\item{\textbf{Gaussian Node Positions:} Node positions drawn from bivariate Gaussian density:
\begin{equation}
f(x_1,x_2) = \frac{1}{\big( 2 \pi \big )^{\frac{d}{2}}} e^{ - \frac{(x_1^2 + x_2^2)}{2}} ,
\end{equation}
on $\mathbb{R}^2$.}
\end{enumerate}
The motivation for considering periodic RGGs is that they are often studied as a means of making the mathematics more analytically tractable by removing boundary effects (for example in \cite{sardellitti_optimal_2012}). Comparing periodic and solid boundaries allows us to understand when boundaries are important in determining network properties. Studying the case of Gaussian RGGs has two motivations. Firstly, it allows us to understand how non-uniformity in the node locations affects the network properties, and secondly, spatial networks with Gaussian distributed node locations have been studied in the case of latent position inference \cite{rastelli_properties_2016} meaning that any results obtained will be practically applicable. 

\begin{figure}[h]
\includegraphics[width=1.0\textwidth]{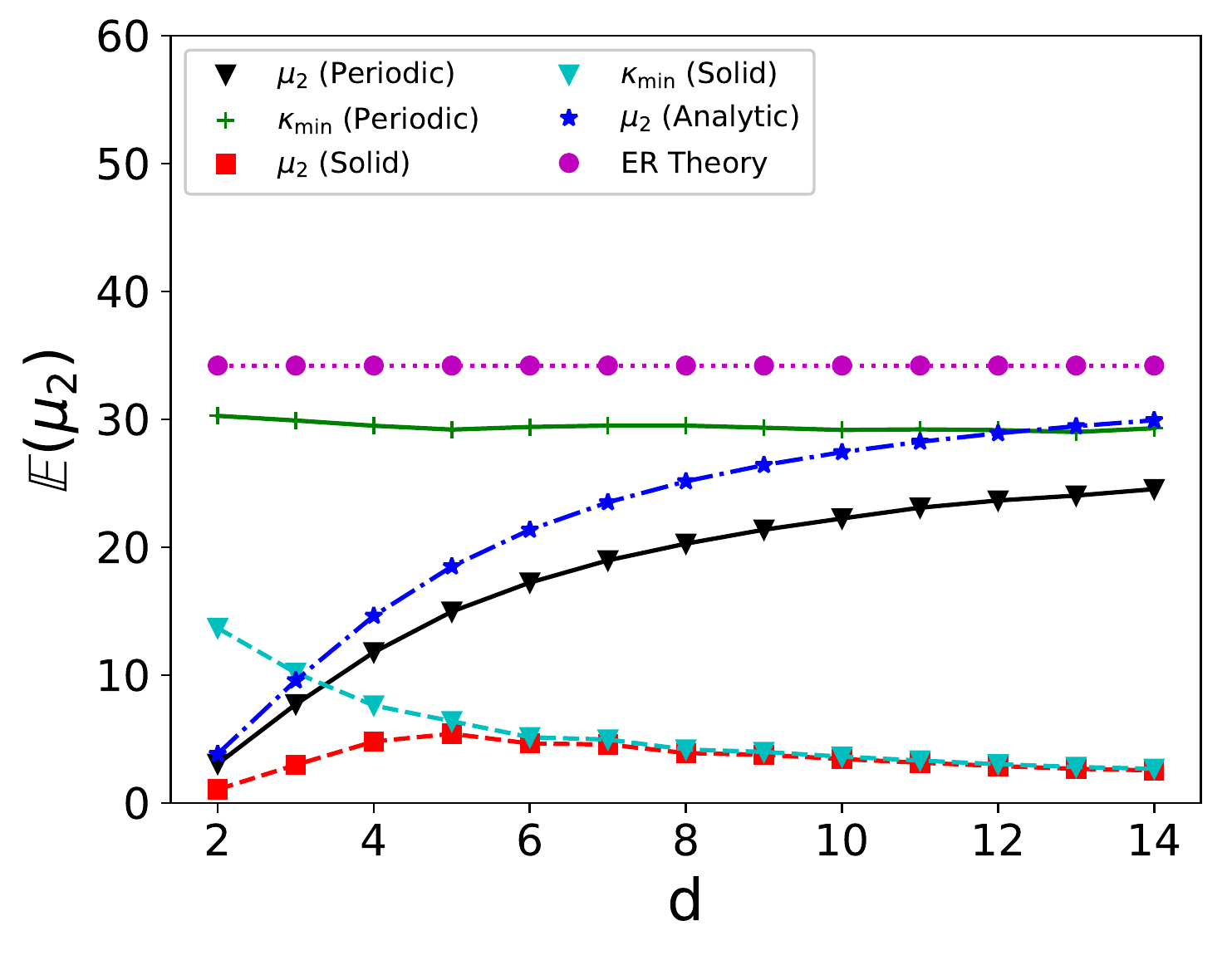}
\vspace*{-1.5em}
\caption{In high dimensional RGGs the behaviour of $\mu_2$ can be accounted for by that of $\kappa_{\mathrm{min}}$. Plot showing the behaviour of $\mathbb{E}(\mu_2)$ for RGGs in $[0,1]^d$ as a function of dimensionality for RGGs with periodic and solid boundaries (black triangles and red squares). Shown for RGGs with $N=10^3$, $\kappa=50$. Also shown are values of $\mathbb{E}(\kappa_{\mathrm{min}})$ for the same ensembles (green crosses and cyan triangles) and theoretical predictions of $\mu_2$ made using \eqref{theory_in_paper} (blue stars). For periodic RGGs the simulated value of $\mathbb{E}(\mu_2)$  as a function of $d$ closely matches that of the theoretical prediction obtained using equation \eqref{theory_in_paper}. For larger values of $d$ both of these values approach the observed value of $\kappa_{\mathrm{min}}$ and appear to be bounded by the theoretical value of $\mathbb{E}(\kappa_{\mathrm{min}})$ for ER graphs (purple circles). For solid RGGs the observed value of $\mathbb{E}(\mu_2)$ (red squares) is well approximated by the observed value of $\mathbb{E}(\kappa_{\mathrm{min}})$ (cyan triangles). Results are averaged over $200$ simulations. Error bars are smaller than the symbol size.}
\label{Mean_And_CV_Prediction}
\end{figure}

\subsection{Behaviour of $\mathbb{E}(\mu_2)$} \label{av_mu2}

The value of $\mathbb{E}(\mu_2)$ is informative about how well connected the typical graph drawn a particular RGG ensemble is. An analytic approximation for $\mathbb{E}(\mu_2)$ is derived in \cite{sardellitti_optimal_2012} for RGGs with node position in $[0,1)^2$ with toroidal boundary conditions. We have extended this approach to determine an approximate expression for $\mu_2$ for the case of general $d$ (see section \ref{Analytic_Derivation} of the SI). The approximation takes the form:
\begin{equation} \label{theory_in_paper}
\mu_2 \approx \kappa - N R^{\frac{d}{2}} J_{\frac{d}{2}} (2 \pi R ) ,
\end{equation}
where $J_{\alpha}$ is a Bessel function of the first kind of order $\alpha$. This approximation holds for the case of $N \rightarrow \infty$, however, as we show below, it provides a good approximation to the behaviour of $\mathbb{E}(\mu_2)$ for the case of finite $N$.

Higher dimensional RGGs have similar properties to Erd\H{o}s-R\'enyi (ER) graphs \cite{dall_random_2002}. In \cite{jamakovic_robustness_2008} they derive the mean and variance of $\mathbb{P}(\mu_2)$ for ER networks by approximating $\mu_2$ by $\kappa_{\mathrm{min}}$. This motivates us to compare the values of $\mu_2$ observed to the ensemble mean value of the minimum degree, $\mathbb{E}(\kappa_{\mathrm{min}})$. Figure \ref{Mean_And_CV_Prediction} shows the behaviour of $\mathbb{E}(\mu_2)$ as a function of $d$. Also shown is the behaviour of $\mathbb{E}(\kappa_{\mathrm{min}})$ for the corresponding graph ensembles as well as the predictions of $\mathbb{E}(\mu_2)$ made based on equation (3) of \cite{jamakovic_robustness_2008} and equation \eqref{theory_in_paper}. 

For RGGs with periodic boundaries, as $d$ is increased for fixed $N$ and $\kappa$, we observe an increase in the value of $\mathbb{E}(\mu_2)$ from an initially small value to one which approaches $\mathbb{E}(\kappa_{\mathrm{min}})$. In this regime, both the analytic approximation of \cite{jamakovic_robustness_2008} and that from equation \eqref{theory_in_paper} can be used to approximate the value of $\mu_2$. For RGGs with solid boundaries the algebraic connectivity is typically much lower than for the periodic case. Despite this, we see that in higher dimensions the behaviour of $\mathbb{E}(\mu_2)$ for RGGs with solid boundaries is also well approximated by the estimated value of $\mathbb{E}(\kappa_{\mathrm{min}})$.

\begin{figure}
\centering
\subfloat[]{\label{hist_d2}\includegraphics[width=.9\textwidth]{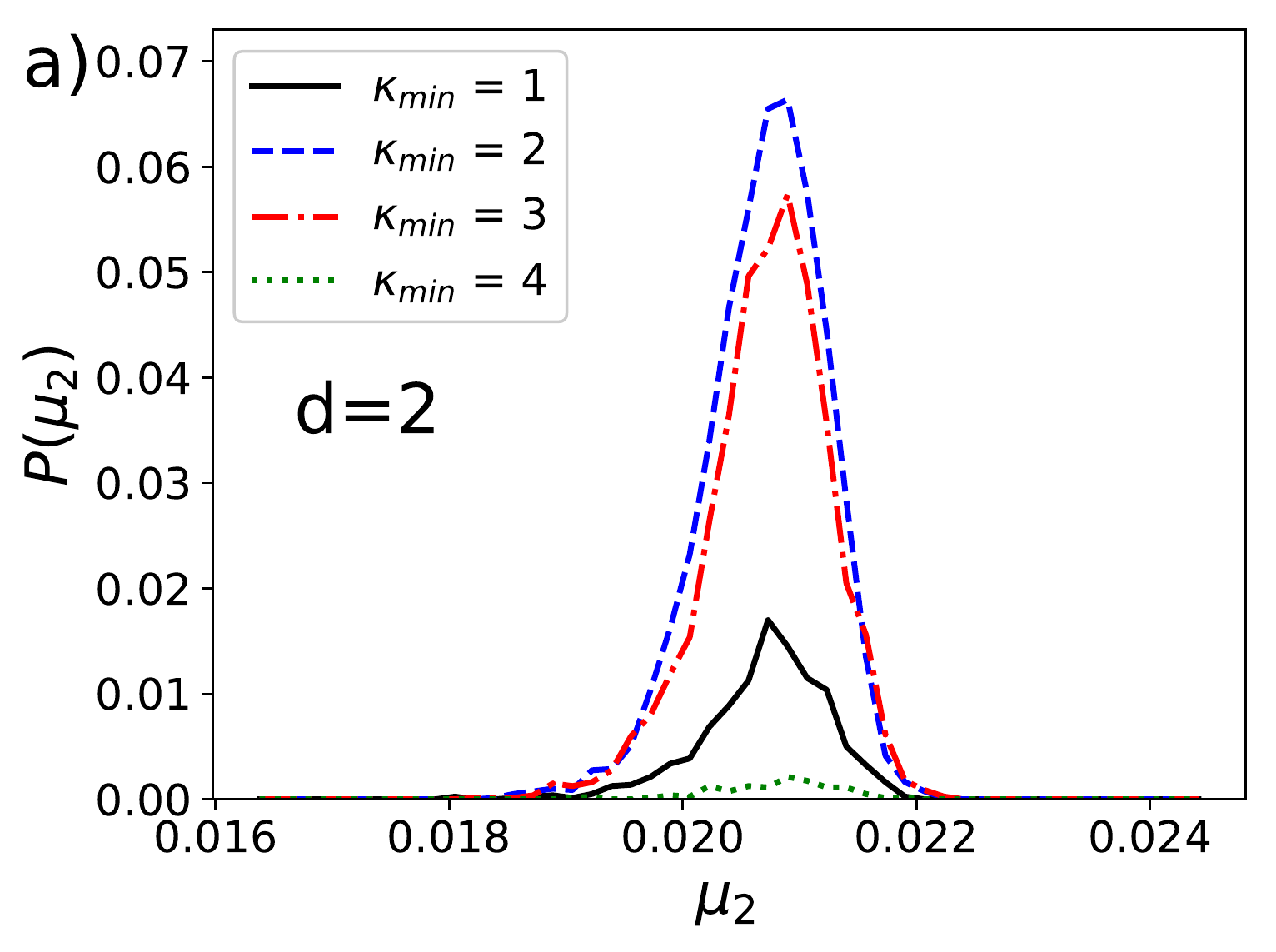}}

\vspace*{-2.0em}

\subfloat[]{\label{hist_d5}\includegraphics[width=.9\textwidth]{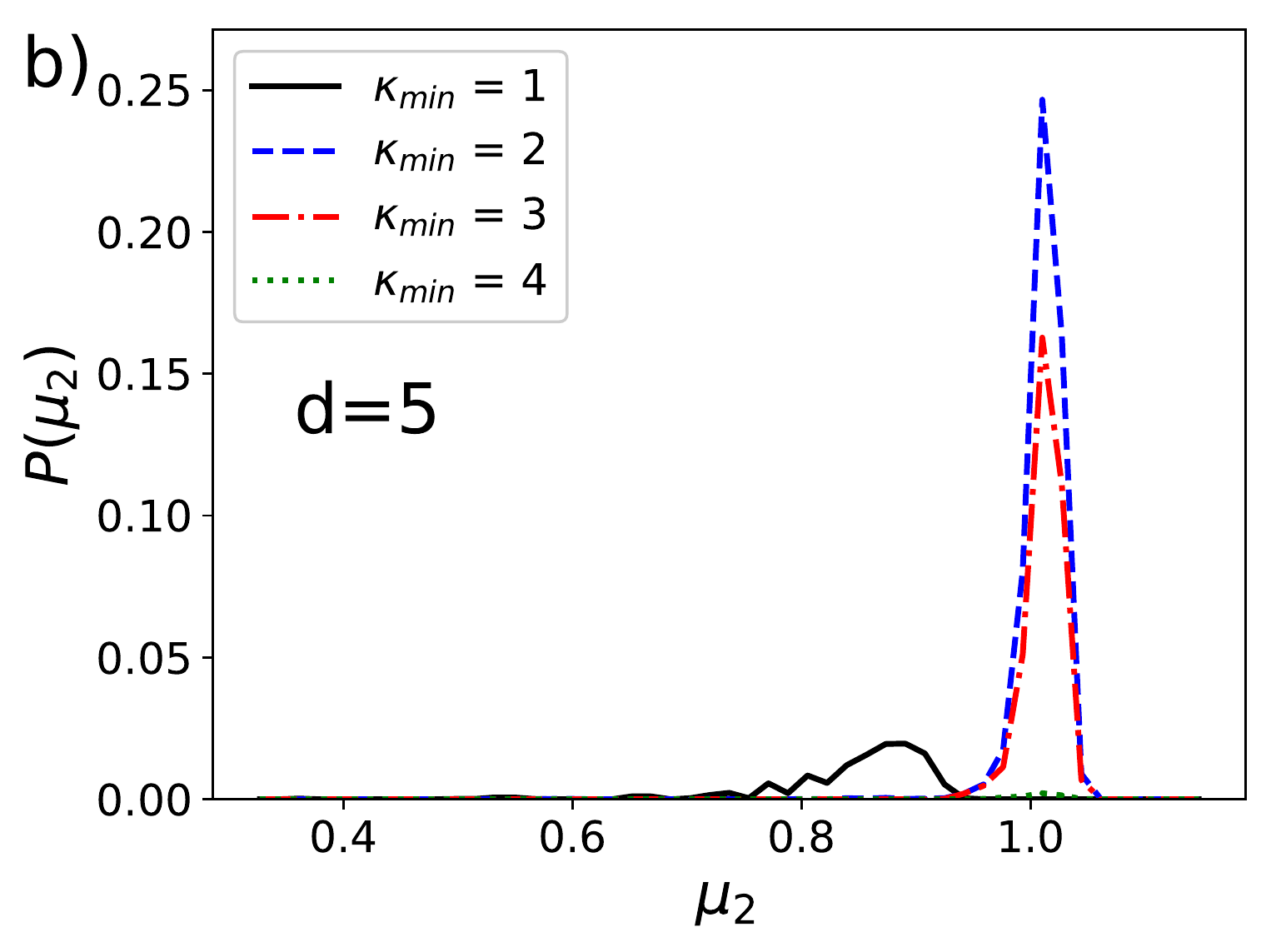}}
\vspace*{-2.0em}

\subfloat[]{\label{hist_d10}\includegraphics[width=.9\textwidth]{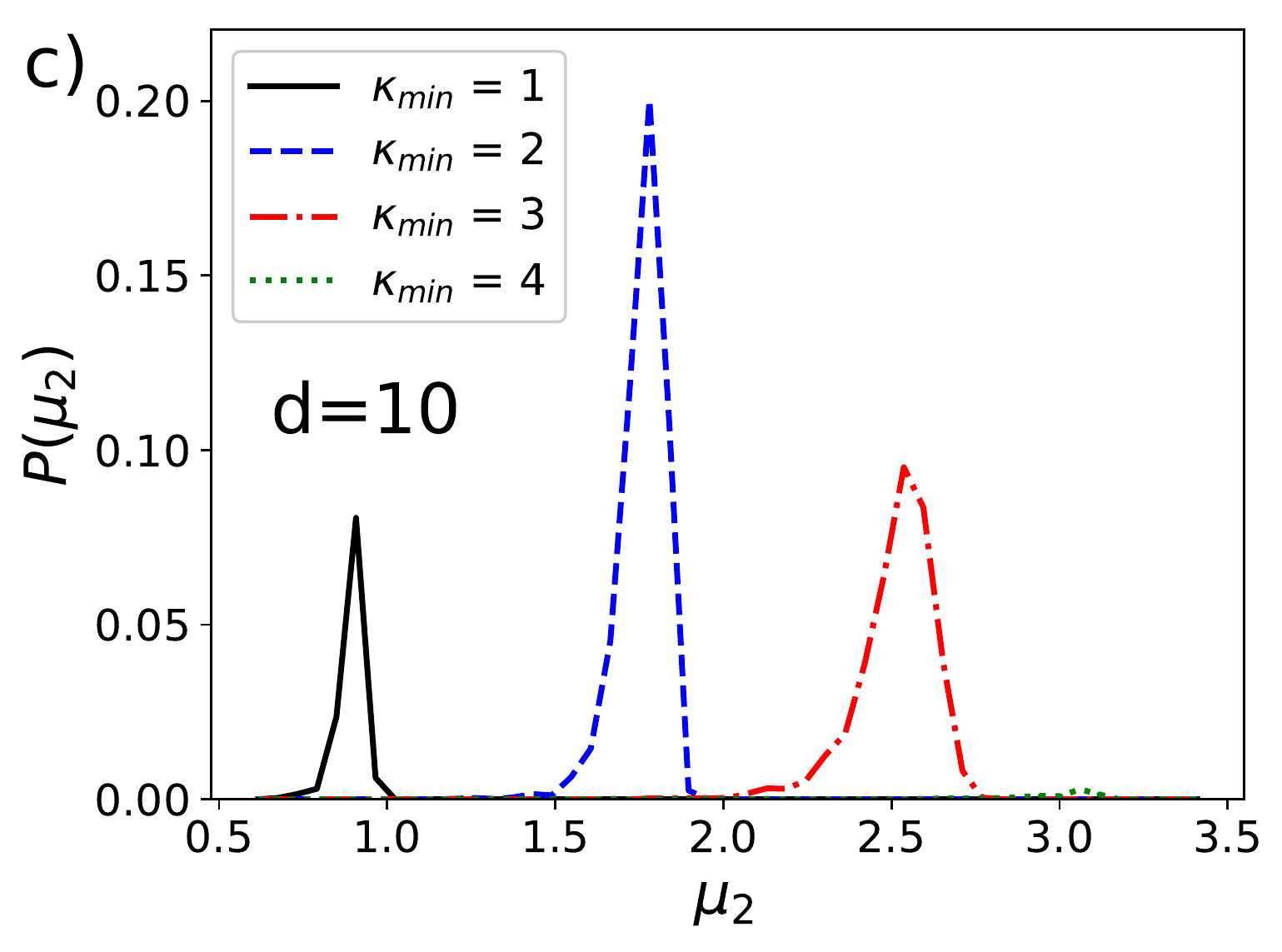}}
\vspace*{-2.0em}

\caption{The modal structure of $\mathbb{P}(\mu_2)$ varies with the spatial dimension. Histograms of $\mathbb{P}(\mu_2)$ with values of $\mu_2$ binned according to the corresponding value of $\kappa_{\mathrm{min}}$ for each graph. Shown for RGG ensembles with $N=10^4$, $C=1.5$ with periodic boundary conditions for a) $d=2$ , b) $d=5$ and c) $d=10$. For 2d RGGs the distribution of $\mu_2$ values consists of a single mode while in the higher dimensional case we observe multiple modes corresponding to graphs with different $\kappa_{\mathrm{min}}$ values. Each plot shows 8000 draws from the corresponding ensemble with the counts normalized by the total number of samples.}
\label{Mu2_Hists}
\end{figure}

\subsection{Effect of Dimensionality on Ensemble Variability} \label{dimvar}

The results shown in Figure \ref{Mean_And_CV_Prediction} demonstrate that the average value of $\mu_2$ can be accounted for by $\kappa_{\mathrm{min}}$ in higher dimensional uniform RGG ensembles. Correlated means do not necessarily imply that the value of $\kappa_{\mathrm{min}}$ for a particular graph is predictive of its $\mu_2$. We nonetheless find that it is possible to account for the fluctuations in $\mu_2$ across the ensemble by considering the fluctuations in $\kappa_{\mathrm{min}}$. Figure \ref{Mu2_Hists} shows histograms of $\mathbb{P}(\mu_2)$ with samples binned according to the corresponding $\kappa_{\mathrm{min}}$ value of the graph. In lower dimensional systems where $CV(\mu_2)$ is small, $\mathbb{P}(\mu_2)$ consists of a single mode (Figure \ref{hist_d2}). For higher dimensional systems we see $\mathbb{P}(\mu_2)$ split into multiple modes (Figure \ref{hist_d10}), each of which is associated with graphs which have a different value of $\kappa_{\mathrm{min}}$. For the case of $d=5$ (Figure \ref{hist_d5}) we observe an intermediate case where graphs with $\kappa_{\mathrm{min}}=1$ have smaller values of $\mu_2$ while graphs with $\kappa_{\mathrm{min}}=2,3,4$ have larger values which fall within the same mode.

\begin{figure*}
\centering

\subfloat[]{\label{CV_as_d_Periodic}\includegraphics[width=.32\textwidth]{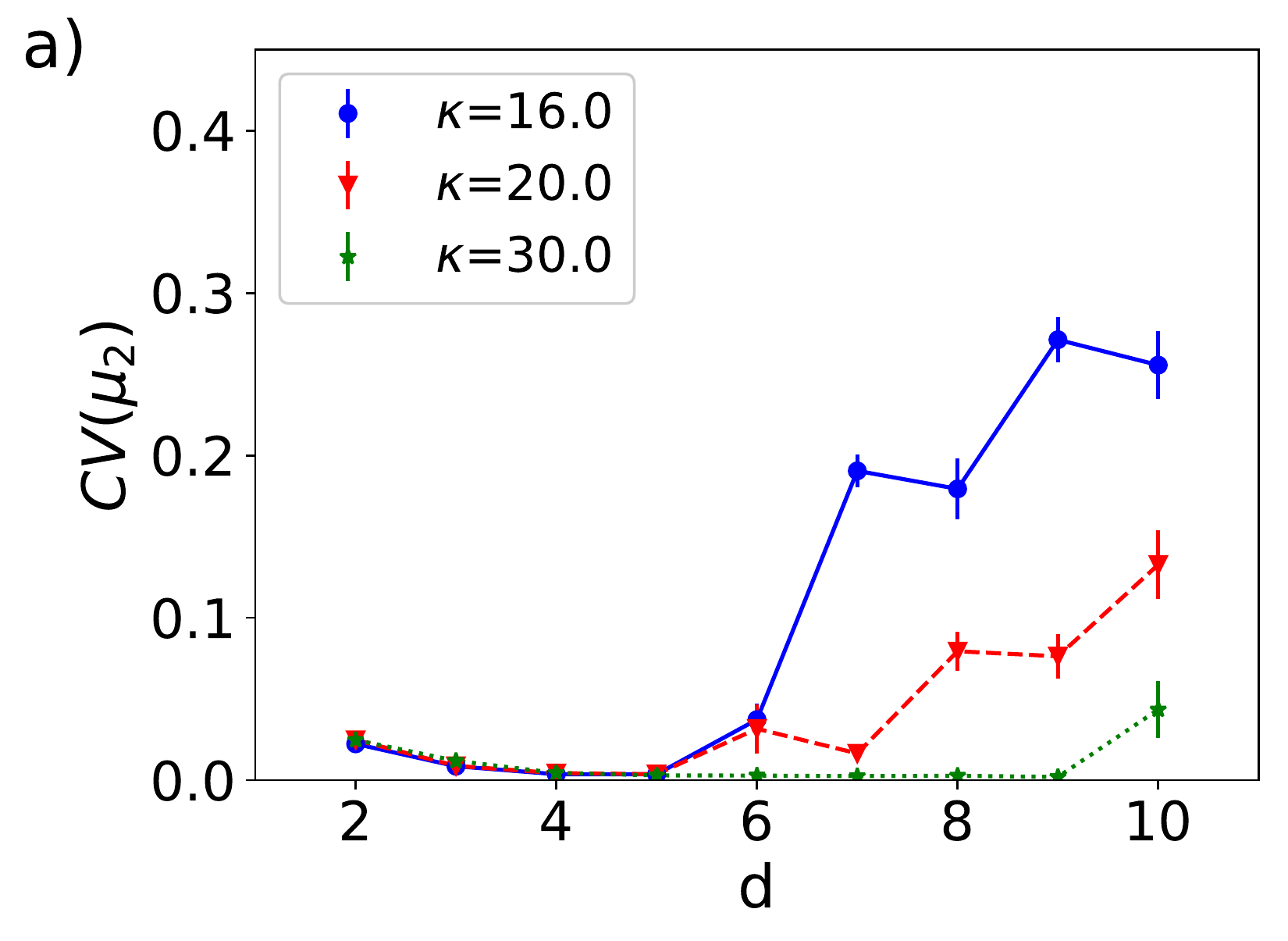}}
\subfloat[]{\label{Corr_Periodic}\includegraphics[width=.32\textwidth]{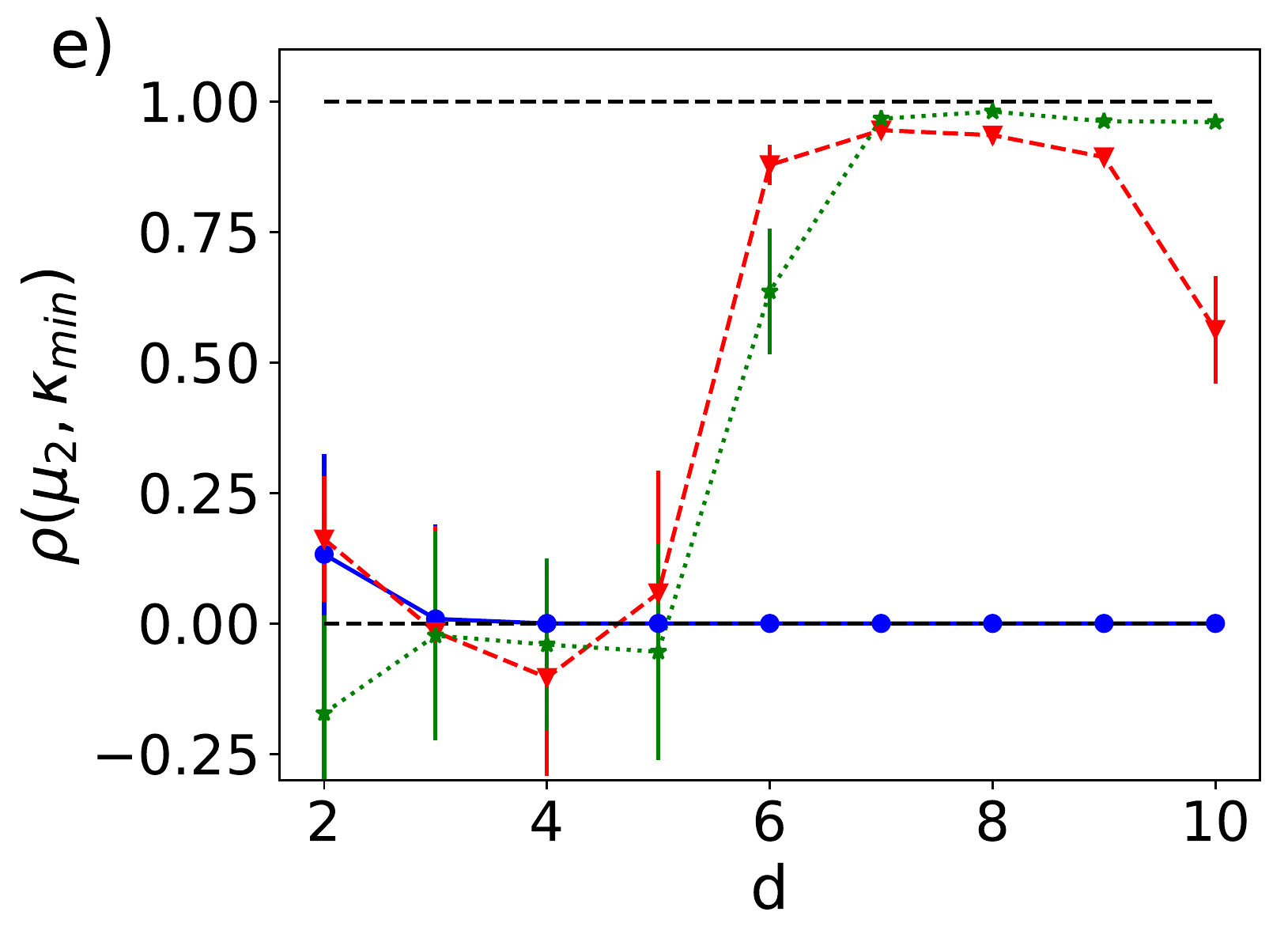}}
\subfloat[]{\label{Kmin_Periodic}\includegraphics[width=.32\textwidth]{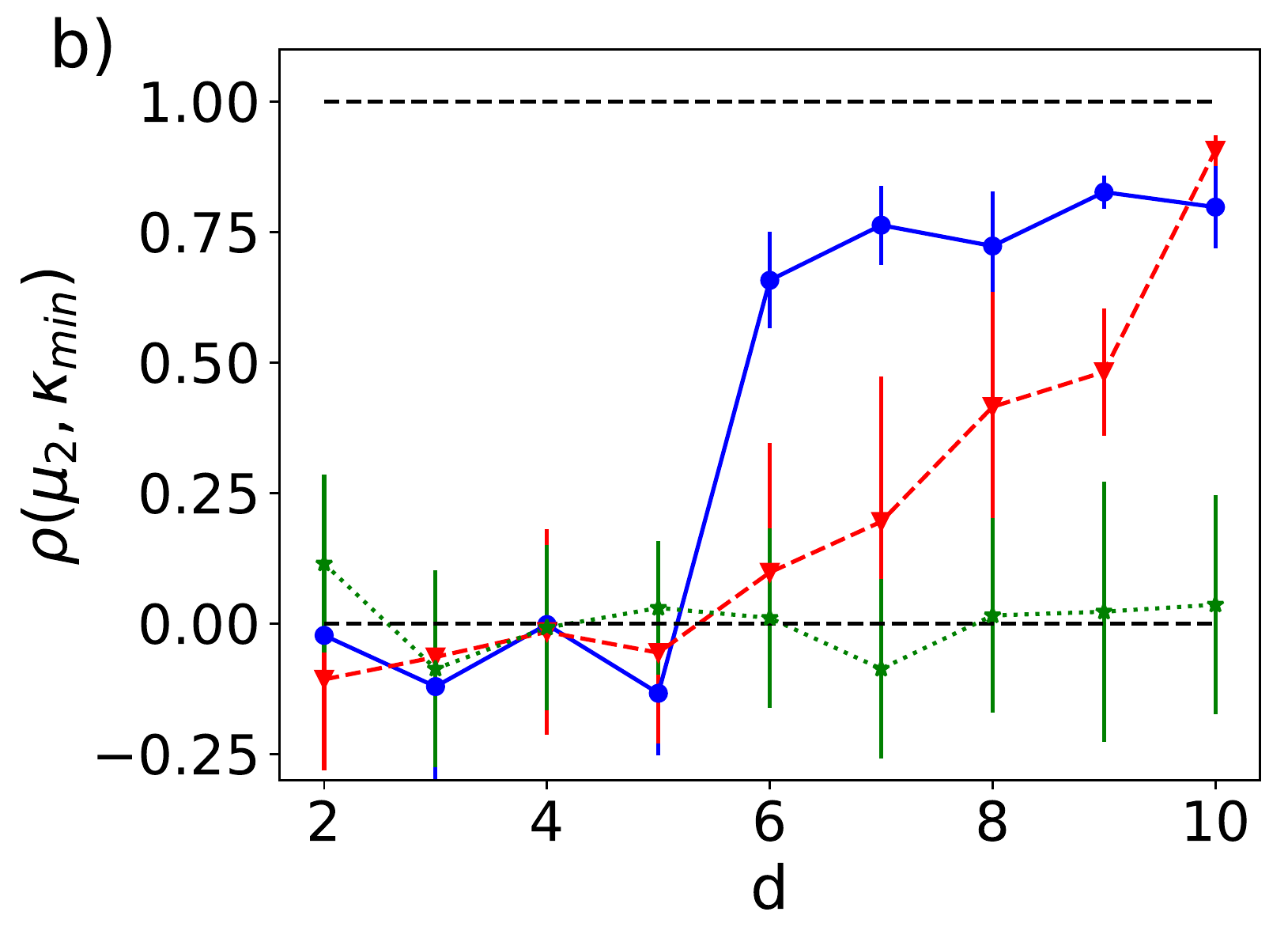}}

\vspace*{-3.0em}
\subfloat[]{\label{CV_as_d_Solid}\includegraphics[width=.32\textwidth]{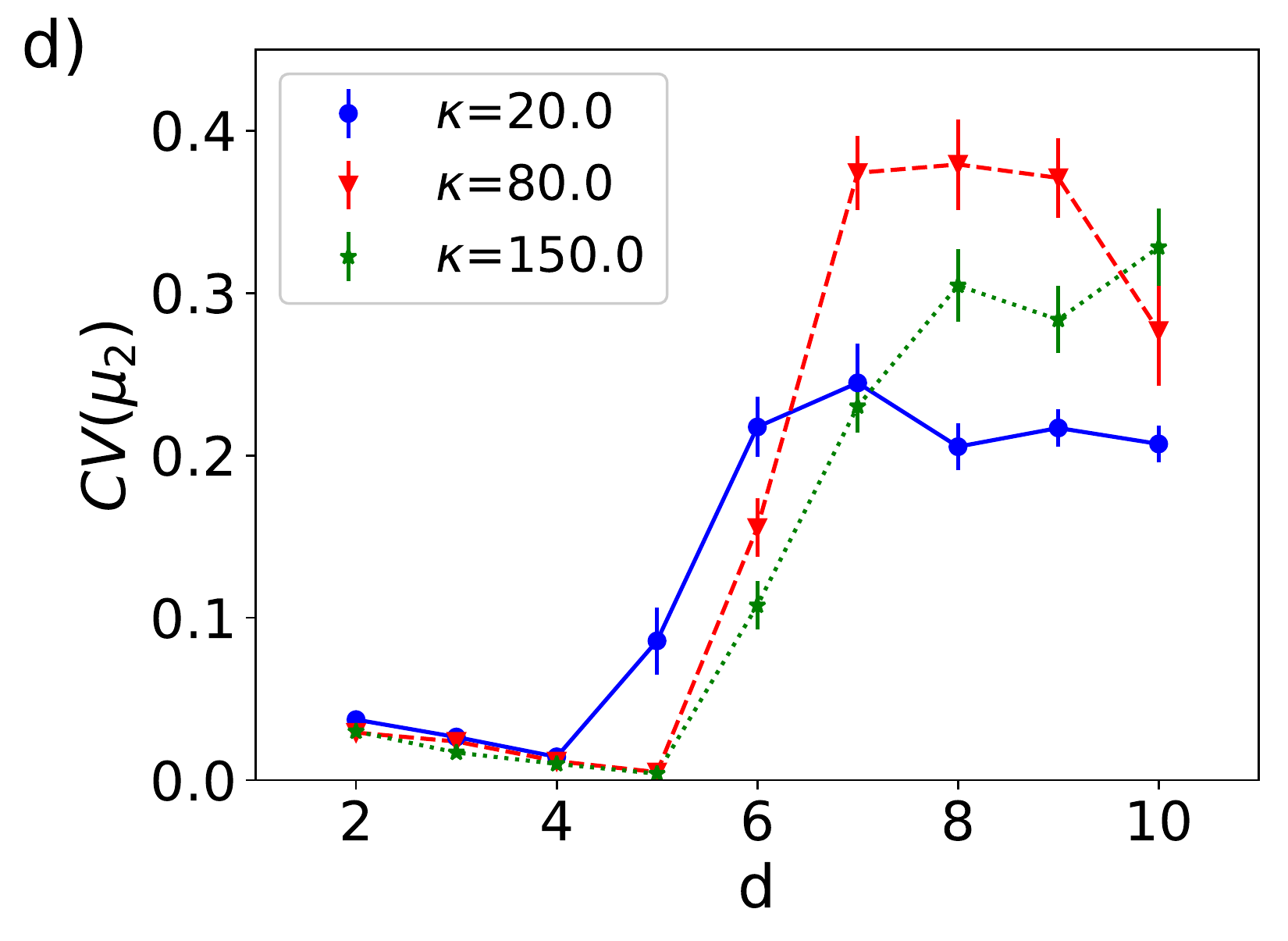}}
\subfloat[]{\label{Corr_Solid}\includegraphics[width=.32\textwidth]{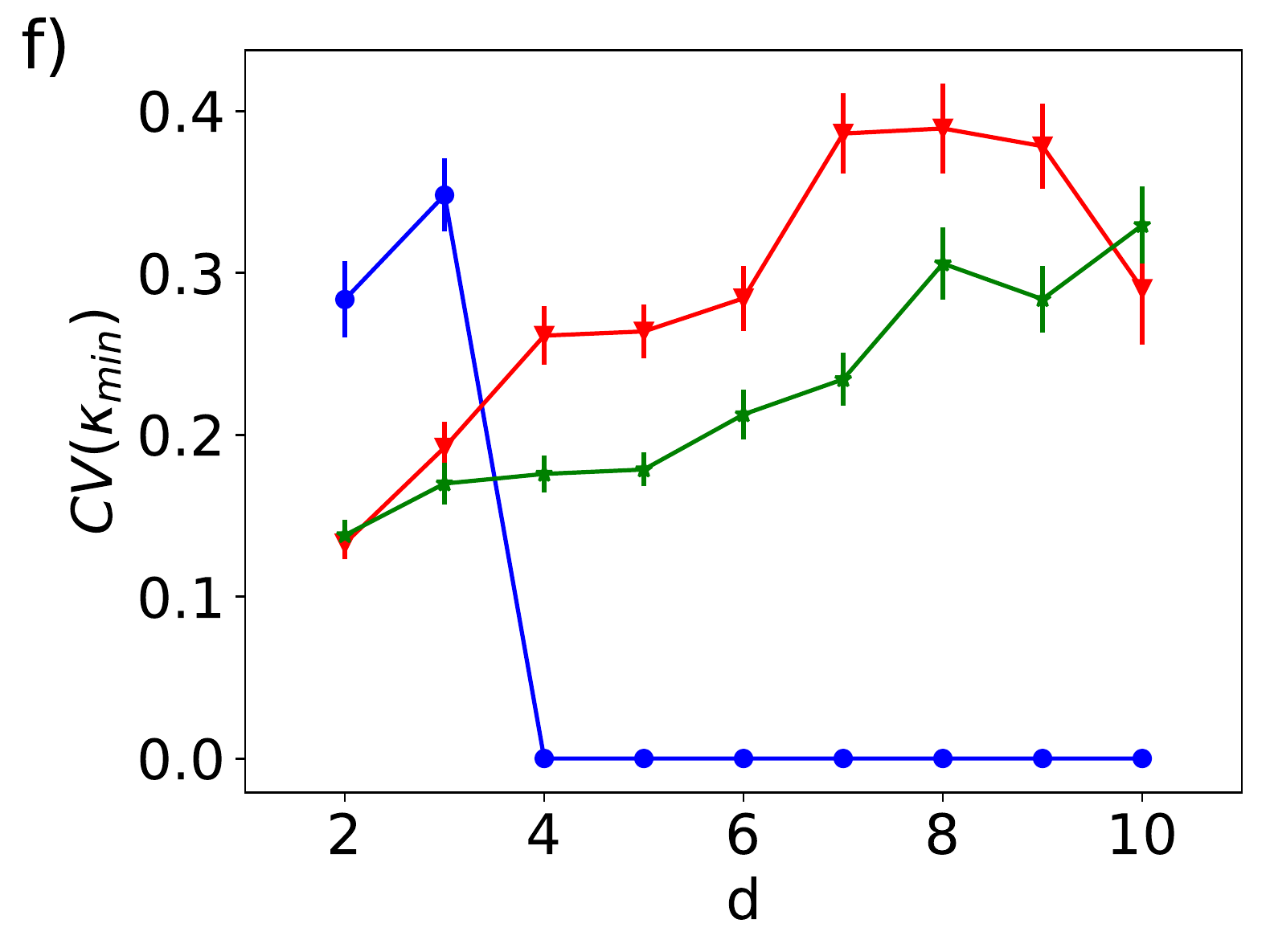}}
\subfloat[]{\label{Kmin_Solid}\includegraphics[width=.32\textwidth]{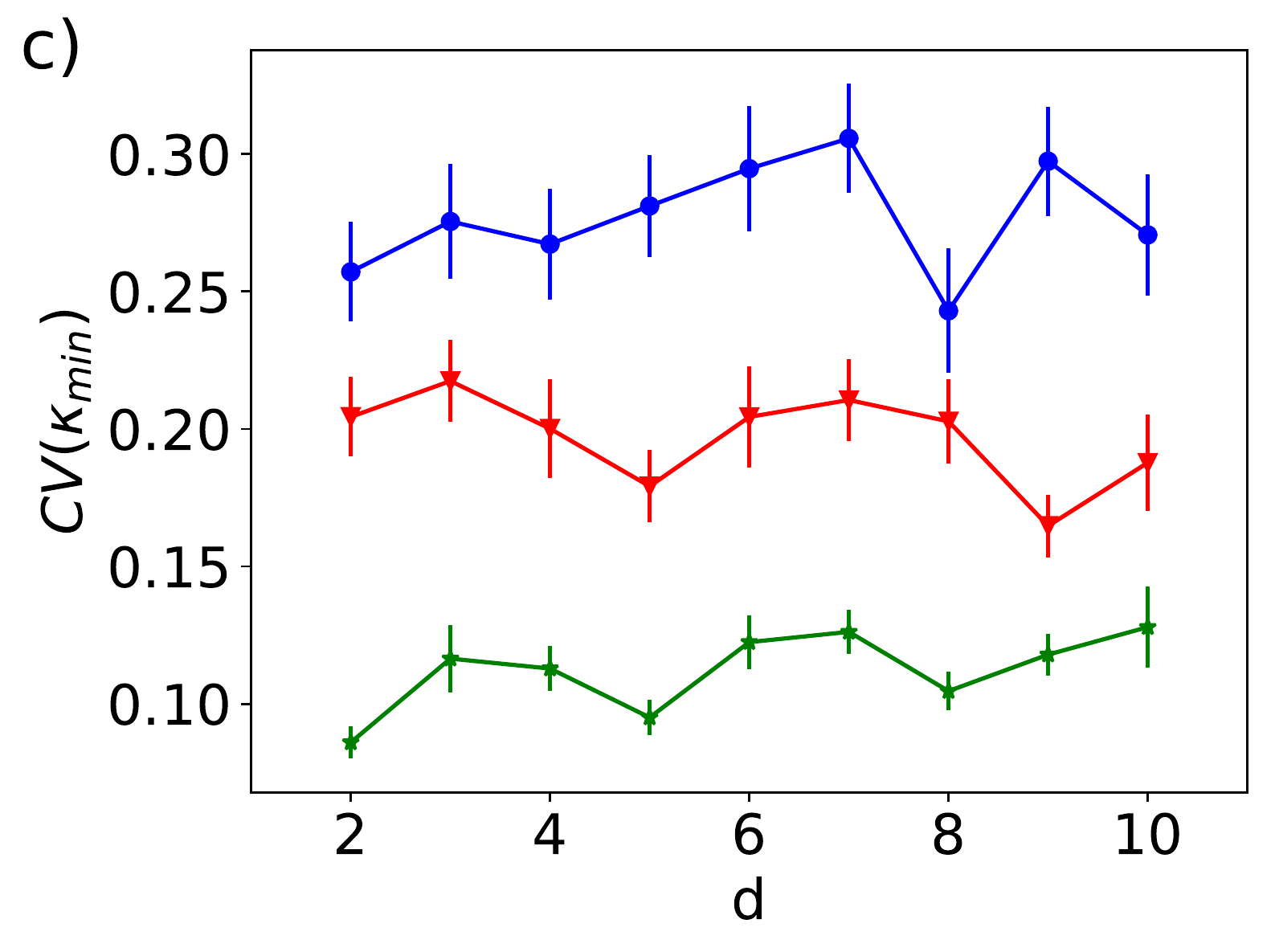}}
\vspace*{-2.0em}
\caption{$CV(\mu_2)$ can be substantial in large RGGs in higher dimensions. a) and d) show the behaviour of $CV(\mu_2)$ as a function of $d$ for RGGs with periodic and solid boundaries respectively. Shown for RGG ensembles with $N=10^5$ for different values of the average degree, $\kappa$. Different values of $\kappa$ are shown for the solid RGG ensembles as a larger value of $\kappa$ is required to achieve full connectivity in this case. b) and e) show the behaviour of $\rho(\mu_2,\kappa_{\mathrm{min}})$ for the same parameter range while c) and f) show the corresponding values of $CV(\kappa_{\mathrm{min}})$. Large values of $CV(\mu_2)$ can be attributed to fluctuations in $\kappa_{\mathrm{min}}$ with the exception of solid RGG ensembles with the low mean degrees ($\kappa=20.0$) for which $\kappa_{\mathrm{min}}=1$ for all graphs in ensembles with $d\geq4$. In this case there is no observable correlation between $\mu_2$ and $\kappa_{\mathrm{min}}$. Error bars on $CV(\mu_2)$ are obtained using the standard error on the coefficient of variation (see section \ref{CV_SE} of the SI) and those on $\rho(\mu_2,\kappa_{\mathrm{min}})$ represent 95\% confidence intervals obtained via bootstrapping. Values of the mean degree reported correspond to input values of the desired mean degree for the algorithm described in section \ref{meandeg_alg} of the SI. Each data point corresponds to 100 samples from the corresponding ensemble.}
\label{fig_CV_as_d}
\end{figure*}

The correlation between $\mu_2$ and $\kappa_{\mathrm{min}}$ can drive large fluctuations in $\mu_2$ for relatively large RGGs ($N=10^5$). Figures \ref{CV_as_d_Periodic} and \ref{CV_as_d_Solid} show the behaviour of $CV(\mu_2)$ as a function of dimension for periodic and solid RGGs. In both cases the value of $CV(\mu_2)$ increases significantly with dimension and can reach values greater than 0.2. The increase in $CV(\mu_2)$ was found to coincide with an increase in $\rho(\mu_2,\kappa_{\mathrm{min}})$ (Figures \ref{Corr_Periodic} and \ref{Corr_Solid}) which demonstrates that the fluctuations in $\kappa_{\mathrm{min}}$ are also the dominant factor driving the fluctuations in $\mu_2$ in these ensembles.

For the case of periodic boundaries, the value of $CV(\mu_2)$ was found to be largest in magnitude for RGG ensembles with low values of $\kappa$. In these graph ensembles $\kappa_{\mathrm{min}}$ typically takes a few relatively small values (e.g 1,2,3) and shows large fluctuations about its average value (Figure \ref{Kmin_Periodic}). For larger $\kappa$ the relative size of the fluctuations in $\kappa_{\mathrm{min}}$ is smaller which leads to less variability in $\mu_2$. As $\kappa$ is decreased towards and below the threshold of full connectivity as $C=1$ the value of $CV(\mu_2)$ will increase further. This increase occurs even in lower dimensional ensembles (see section \ref{Percol_Sect} of the SI). In the case where we are below the threshold of full connectivity the size of the LCC will fluctuate introducing an additional source of randomness. However, it is worth noting that the value of $CV(\mu_2)$ begins to increase before this suggesting that ensembles consisting of sparser graphs tend to show more variability in their dynamical properties. 

The same trends as above hold for solid RGGs, apart from in the case of the lowest value of $\kappa$ studied. In this case the RGG ensembles are such that we typically always have $\kappa_{\mathrm{min}}=1$ and there will not be any observable correlation between $\mu_2$ and $\kappa_{\mathrm{min}}$. Consequently, the relatively large values of $CV(\mu_2)$ ($>$0.2) cannot be attributed to variability in $\kappa_{\mathrm{min}}$. In moderate dimensional RGGs with solid boundaries it is possible to generate weakly connected subgraphs and chains in regions of low density in the corners and edges of the domain. The presence (or lack of presence) of these subgraphs can lead to variability in $\mu_2$. We demonstrate that this effect leads to relatively large fluctuations in $\mu_2$ for Gaussian RGGs in section \ref{Fied_Sect} below. We also observe that $CV(\mu_2)$ slightly decreases for values of $d$ greater than 7. We conjecture that this occurs as we go from a regime where the presence of weakly connected subgraphs or chains (as in section \ref{Fied_Sect}) control the behaviour of $\mu_2$ to one where the dominant factor controlling $\mu_2$ is $\kappa_{\mathrm{min}}$ which does not vary in high dimensional solid RGGs with relatively low mean degrees.

\subsection{Effect of Non-uniformity on Ensemble Variability.} \label{nonuniform}

In the previous section we explored the factors which can drive fluctuations in $\mu_2$ for ensembles of RGGs where the distribution of points is uniform. In this section we explore how non-uniformity in the node distribution can influence the statistical properties of $\mu_2$. We study the properties of $\mathbb{P}(\mu_2)$ for the ensemble of Gaussian RGGs in $d=2$ (defined at the start of section \ref{EnsAverage}).

For uniform RGGs we can scale the degree logarithmically in system size (ie. $\kappa=C \log(N)$) in order to tune the connectivity of the system. In RGGs with Gaussian node locations there is no longer a sharp transition to full connectivity at a specific value of $\kappa$ \cite{barnett_spatially_2007}. Instead, we observe a smooth increase in $F_{LCC}$ as $\kappa$ is increased. As a consequence of the above we will report $F_{LCC}$ in order to give an indication of the typical LCC sizes of the ensembles studied. We have chosen values of $\kappa$ so that the majority of ensembles studied have $F_{LCC} \approx 1$. 

Figure \ref{Gauss_CV} illustrates the behaviour of $CV(\mu_2)$ as a function of system size for ensembles of Gaussian RGGs with different $\kappa$. $CV(\mu_2)$ takes consistently large values ($>0.3$) across the range of $N$ and $\kappa$ studied. In the previous section we explained the larger values of $CV(\mu_2)$ by considering the correlation of $\mu_2$ and $\kappa_{\mathrm{min}}$. For the majority of $\kappa$ and $N$ values considered Gaussian RGGs have LCCs with $\kappa_{\mathrm{min}}=1$. This means that the variability in $\mu_2$ cannot be attributed to fluctuations in $\kappa_{\mathrm{min}}$. In section \ref{Fied_Sect} below we will demonstrate how large values of $CV(\mu_2)$ in Gaussian RGG ensembles can be explained by the presence of weakly connected subgraphs occurring at the edge of these systems.

We also observe that the value of $CV(\mu_2)$ increases with $\kappa$ which contrasts with the results observed in homogeneous systems. Our explanation of this is as follows: for Gaussian RGGs which are not quite `fully connected' increasing $\kappa$ (or $R$) increases the area of the region containing nodes connected to the LCC. This effectively increases the `surface area' of the low node density region in which weakly connected subgraphs can exist thus increasing the probability of various rare events which might alter the $\mu_2$ value of the RGGs. Furthermore, for the case where $\kappa$ is relatively large in smaller graphs (e.g $\kappa=500$ for graphs with $N\lessapprox 5 \times 10^3$) it is possible to generate Gaussian RGGs for which the the value of $\kappa_{\mathrm{min}}$ fluctuates between ensemble members. In this case we observe values of $\rho(\mu_2,\kappa_{\mathrm{min}})$ close to one (See section \ref{Gauss_Min_Fluct} of the SI). This indicates that the value of $\kappa_{\mathrm{min}}$ can also determine that of $\mu_2$ in low dimensional RGGs if the node density is inhomogeneous.

\subsection{Ensemble Variability in the Fiedler Partition.} \label{Fied_Sect}

In order to understand more about the factors which influence the value of $\mu_2$ in Gaussian RGG ensembles we study its corresponding eigenvector, $\underline{u}_2$. This eigenvector is known as the \emph{Fiedler vector} and has application in community detection algorithms \cite{newman_finding_2006}. Studying the components of $\underline{u}_2$ allows us to identify instances of tightly knitted communities in the network. 

We denote the components of $\underline{u}_2$ as $(u_2^1,u_2^2,...,u_2^N)$. A simple heuristic for partitioning the nodes in the network into two groups is to split them according to the sign of the corresponding element of $\underline{u}_2$ \cite{newman_finding_2006}, (also see \cite{van_mieghem_graph_2010} pg 89). That is, we assign node $i$ to group $1$ if $u_2^i \geq 0$ and otherwise we assign it to group 2. 

We also note that for the graph Laplacian, the first eigenvector $\underline{u}_1 = (1,1,...,1)$ will be orthogonal to $\underline{u}_2$. From this it follows that:
\begin{equation}
\sum_{i=1}^N u_2^i = 0 . 
\end{equation}
Consequently, if a subset of the components of $\underline{u}_2$ are much larger in magnitude than the majority, then one of the two groups will be smaller in size. This indicates the presence of a sub-community which is only weakly connected to the majority of the nodes. 

We can keep track of the proportion of nodes assigned to each partition. If we let $N_1$ and $N_2$ be the number of nodes assigned to groups 1 and 2 respectively then we can define:
\begin{equation}
F_P = \frac{\min(N_1,N_2)}{N_{LCC}} . 
\end{equation}
This quantity keeps track of the fraction of nodes assigned to the smaller group. Given that $\sum_i u_i = 0$ a smaller value of $F_P$ will also imply localisation of a large amount of mass of the eigenvector onto a smaller number of nodes. 

\begin{figure}
\centering
\subfloat[]{\label{Gauss_CV}\includegraphics[width=.95\textwidth]{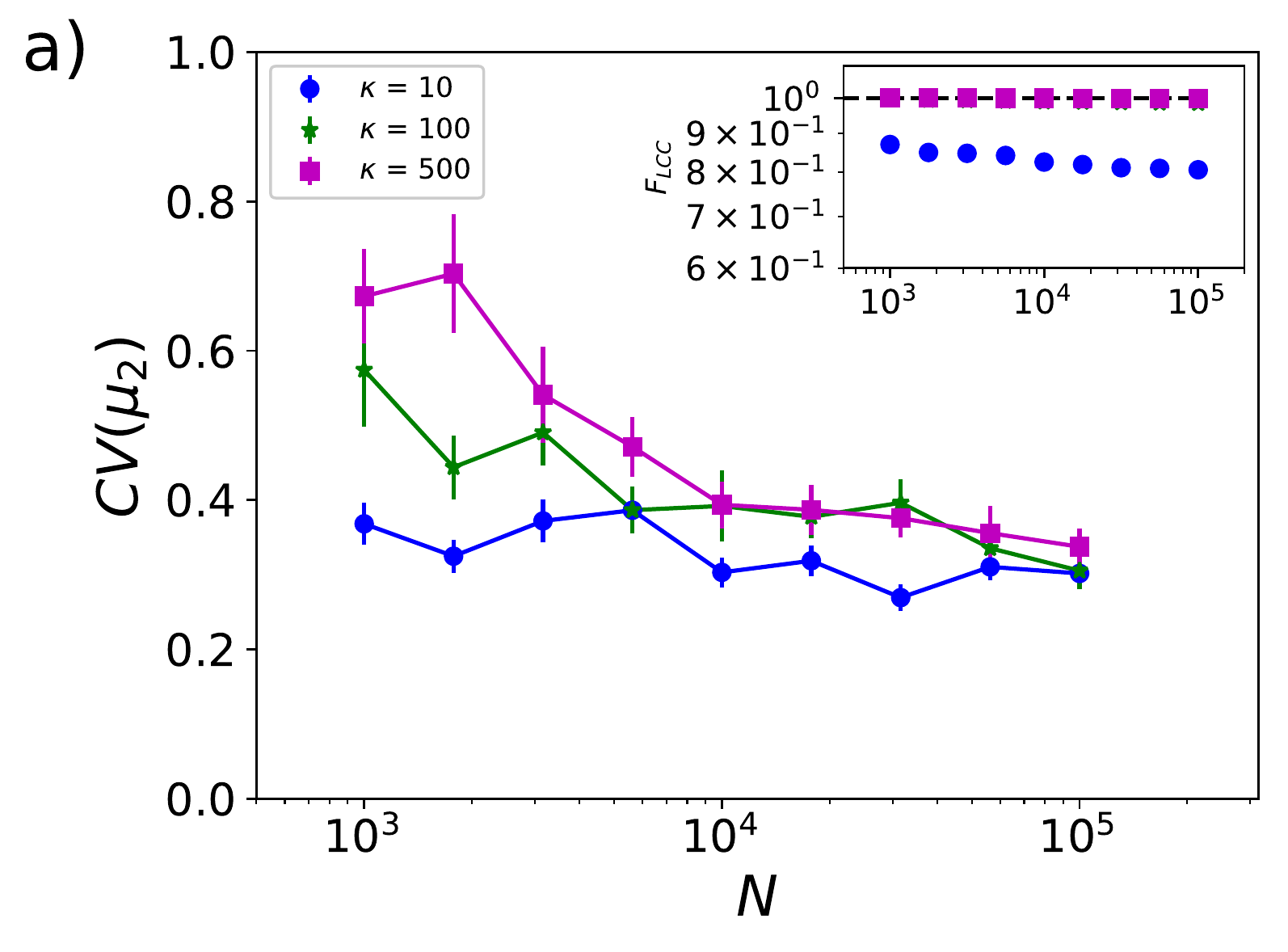}}
\vspace*{-3.0em}

\subfloat[]{\label{Gauss_corr_fp}\includegraphics[width=.95\textwidth]{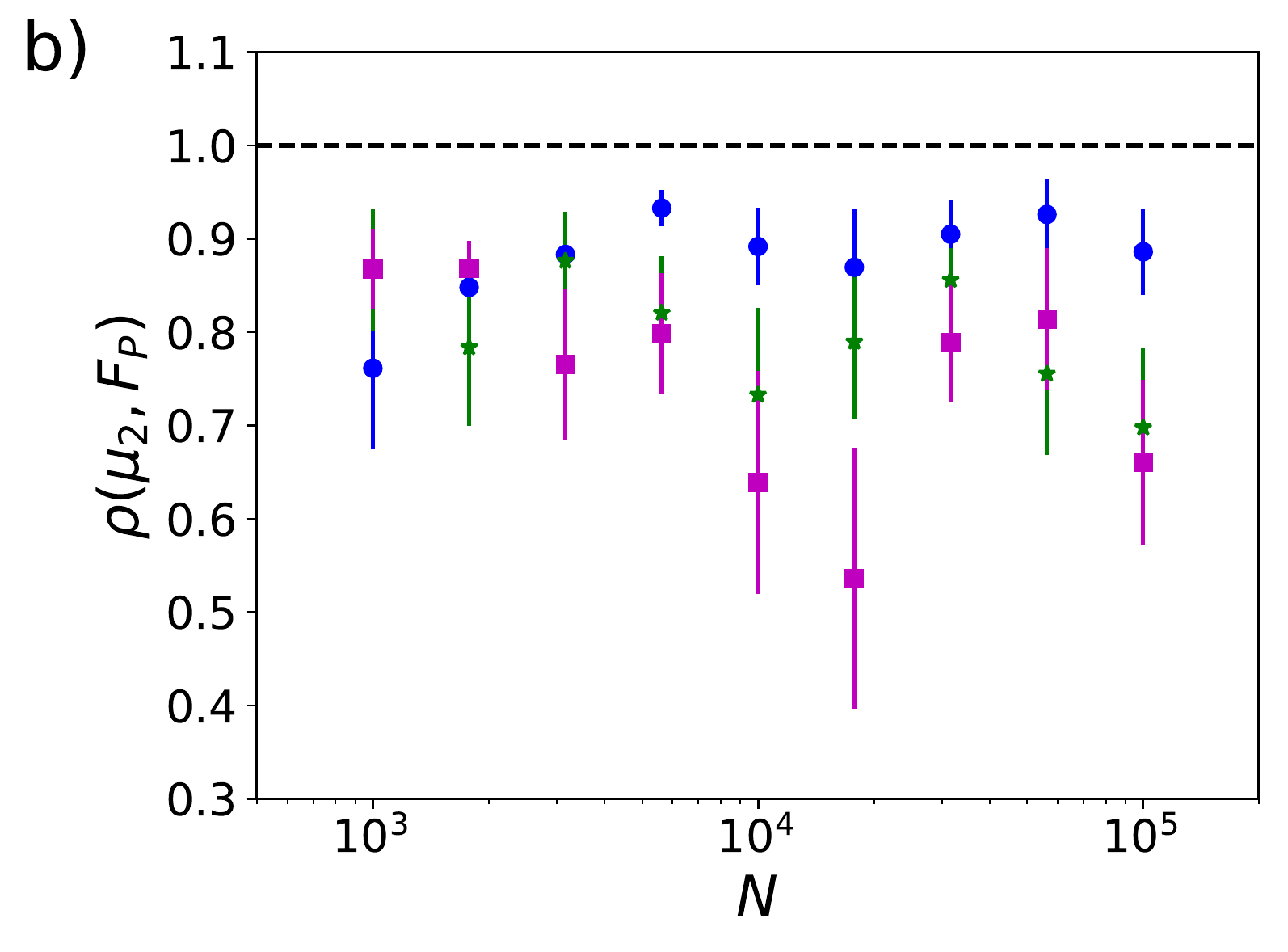}}

\vspace*{-2.0em}
\subfloat[$\mu_2 = 0.01$ , $F_P = 0.03$]{\label{low_mu2_gauss}\includegraphics[width=.5\textwidth]{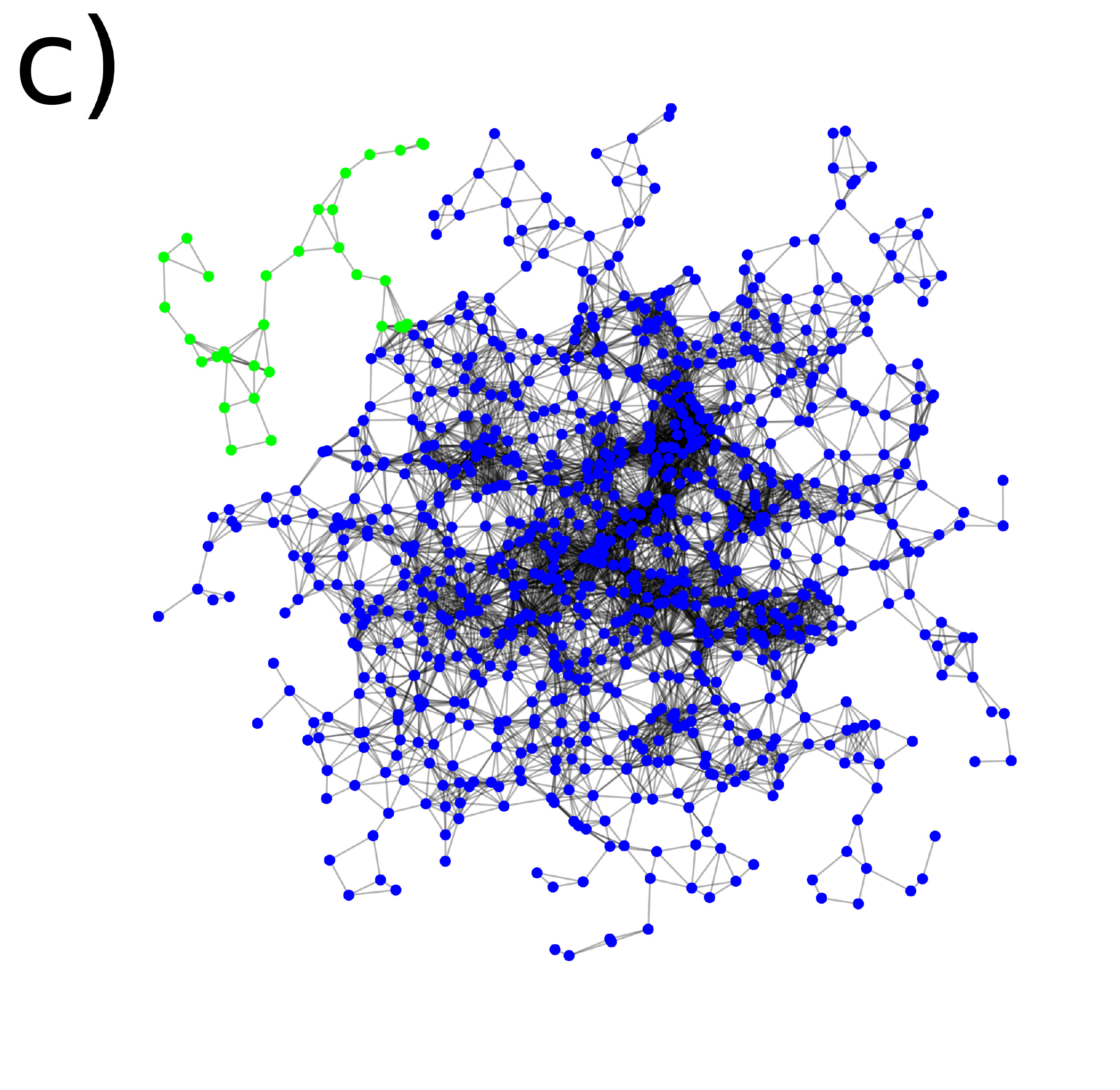}}
\subfloat[$\mu_2 = 0.14$ , $F_P = 0.46$]{\label{high_mu2_gauss}\includegraphics[width=.5\textwidth]{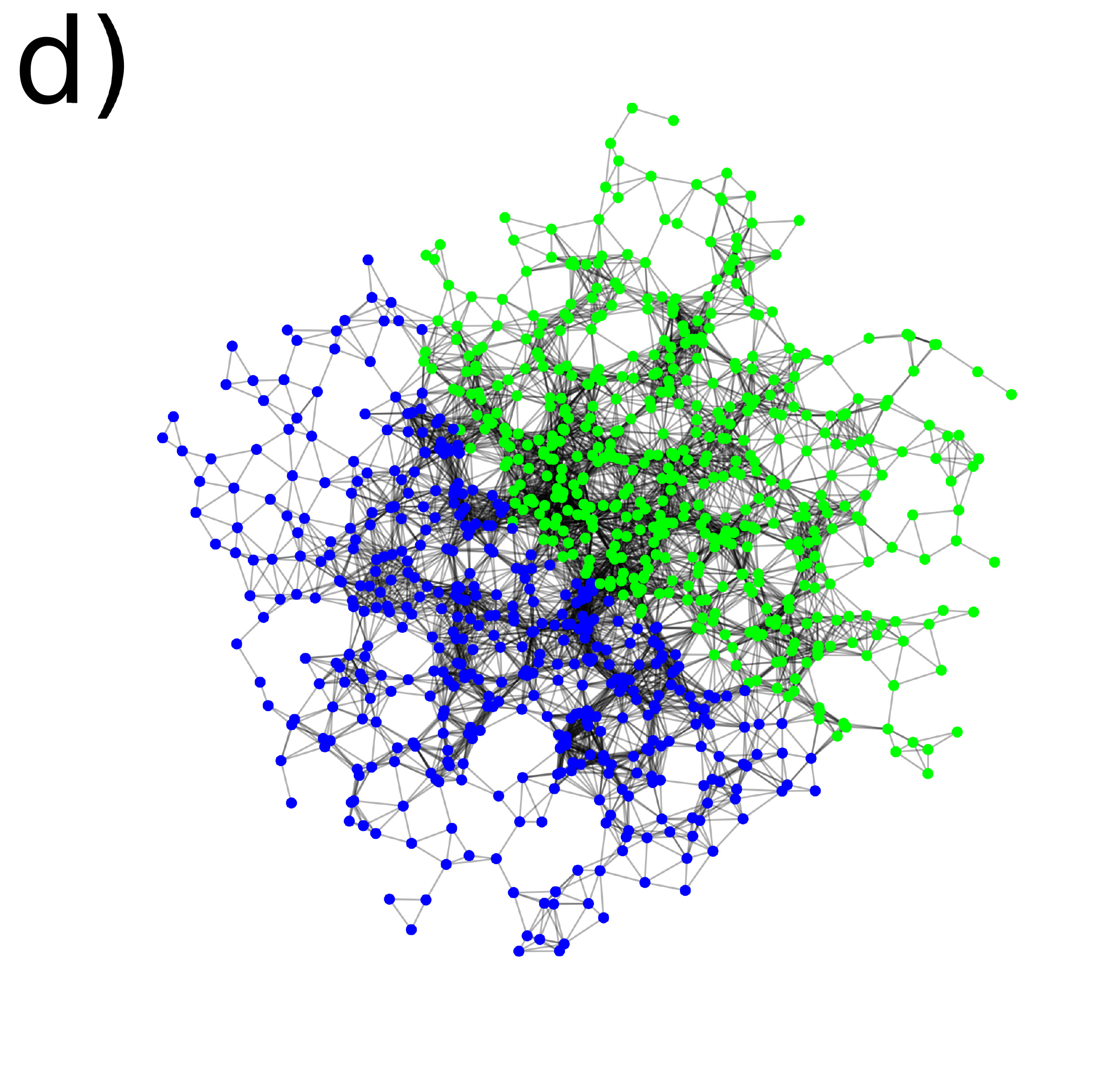}}
\vspace*{-0.5em}
\caption{$CV(\mu_2)$ is large in Gaussian RGG ensembles for a wide range of mean degrees and system sizes. a) Plot showing $CV(\mu_2)$ as a function of system size for Gaussian RGGs with different mean degrees. The inset shows the fraction of nodes in the LCC for the different RGG ensembles. b) Plot showing $\rho(\mu_2,F_P)$ for the same range of parameters as in a). The value of $\rho(\mu_2,F_P)$ remains significant across the range of parameters studied indicating that fluctuations in $\mu_2$ coincide with those in $F_P$. Each data point corresponds to 100 samples from the corresponding ensemble. Figures c) and d) show two graphs drawn from the ensemble of Gaussian RGGs with $N=10^3$ and $\kappa=20$ with the minimum and maximum $\mu_2$ values from 100 draws. This ensemble has $\mathbb{E}(\mu_2)=0.06$. Nodes are coloured according to the sign of the corresponding element of the Fiedler vector. This illustrates that the number of nodes in the Fiedler partition can differ significantly for graphs drawn from the same ensemble. Error bars on $CV(\mu_2)$ and $\rho(\mu_2,F_P)$ were computed using the same methodology as that described for Figure \ref{fig_CV_as_d}. }
\label{Gaussian_Results}
\end{figure}

Figure \ref{Gauss_corr_fp} illustrates the behaviour of $\rho(\mu_2,F_P)$ for the same range of parameters studied in \ref{Gauss_CV}. We observe that this correlation is typically large across the range of $N$ and $\kappa$ values considered. For this range of parameters it was also found that $CV(F_P)$ is typically large. For example, using 100 samples for $N=10^5$, $CV(F_P)$ was estimated to be: 0.47$\pm$0.03,0.89$\pm$0.29 and 1.71$\pm$0.75 for $\kappa=50,100$ and $500$ respectively. This indicates that the number of nodes assigned to the smaller partition is highly variable in Gaussian RGG ensembles. We illustrate this effect for smaller graphs in Figures \ref{low_mu2_gauss} and \ref{high_mu2_gauss} which show draws from the ensemble of Gaussian RGGs with $N=10^3$ and $\kappa=20.0$ with the smallest and largest values of $\mu_2$ respectively. In the former case the smaller partition is associated with a weakly connected subgraph at the edge of the support of the distribution, while in the latter case the partition is much closer to 50/50. We expect that this effect can account for the large values of $CV(\mu_2)$ in large Gaussian RGGs as low node density regimes at the edge of the support of the distribution will persist regardless of system size.

\begin{figure}[]
\centering
\hspace{-10mm}
\subfloat[]{\label{Loc1}\includegraphics[width=.37\textwidth]{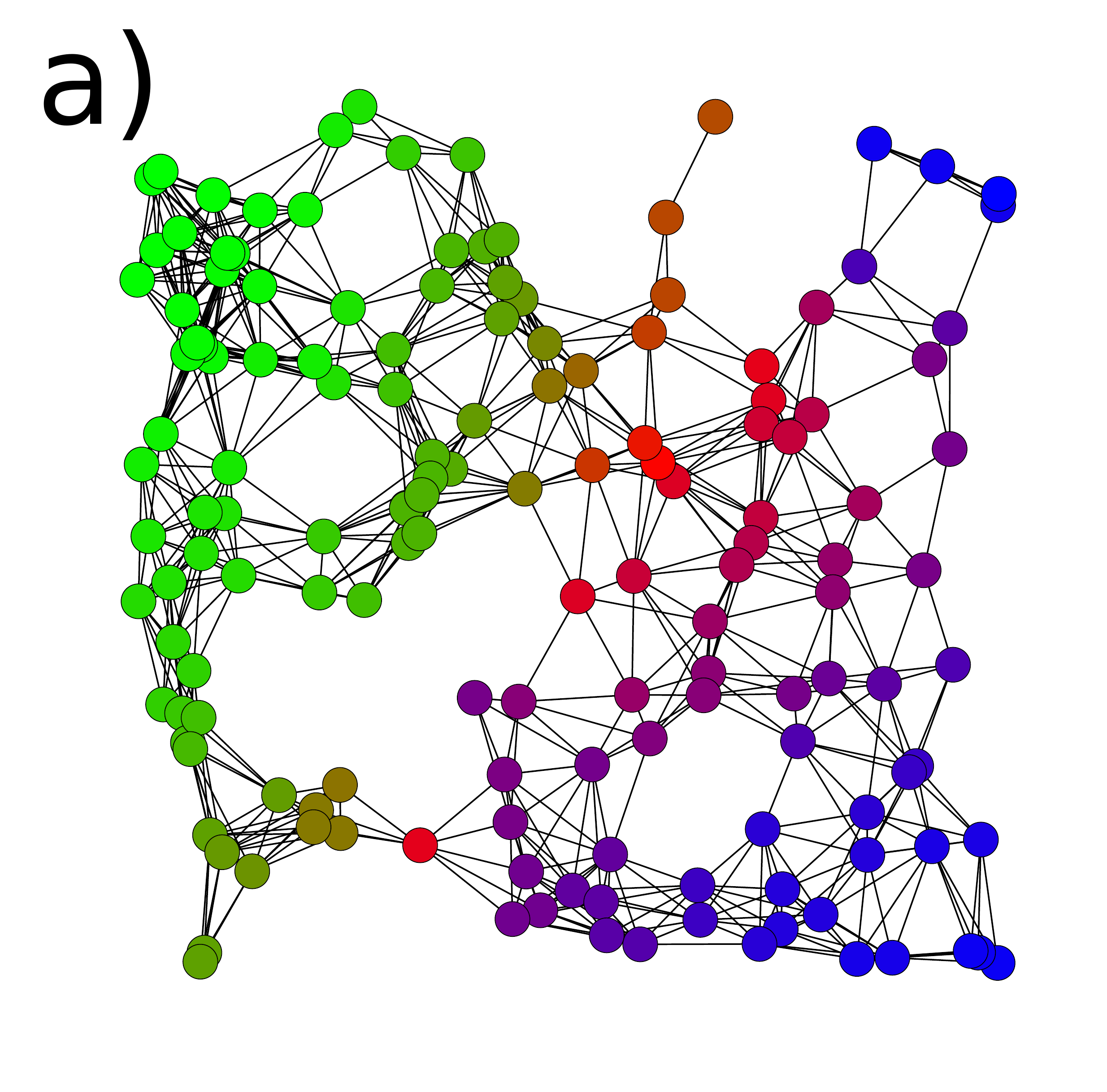}}
\subfloat[]{\label{Loc2}\includegraphics[width=.37\textwidth]{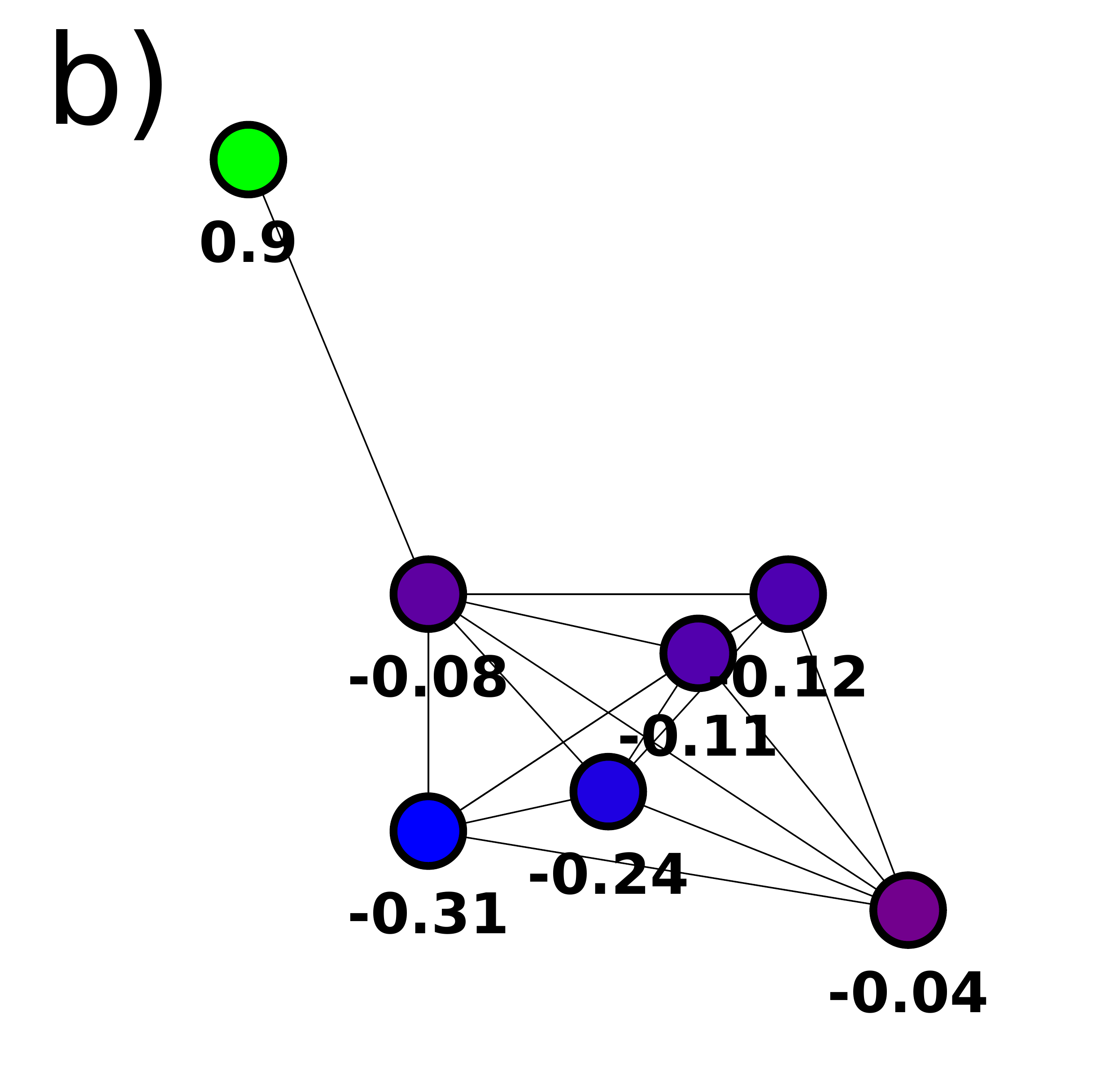}}
\subfloat[]{\label{Loc3}\includegraphics[width=.37\textwidth]{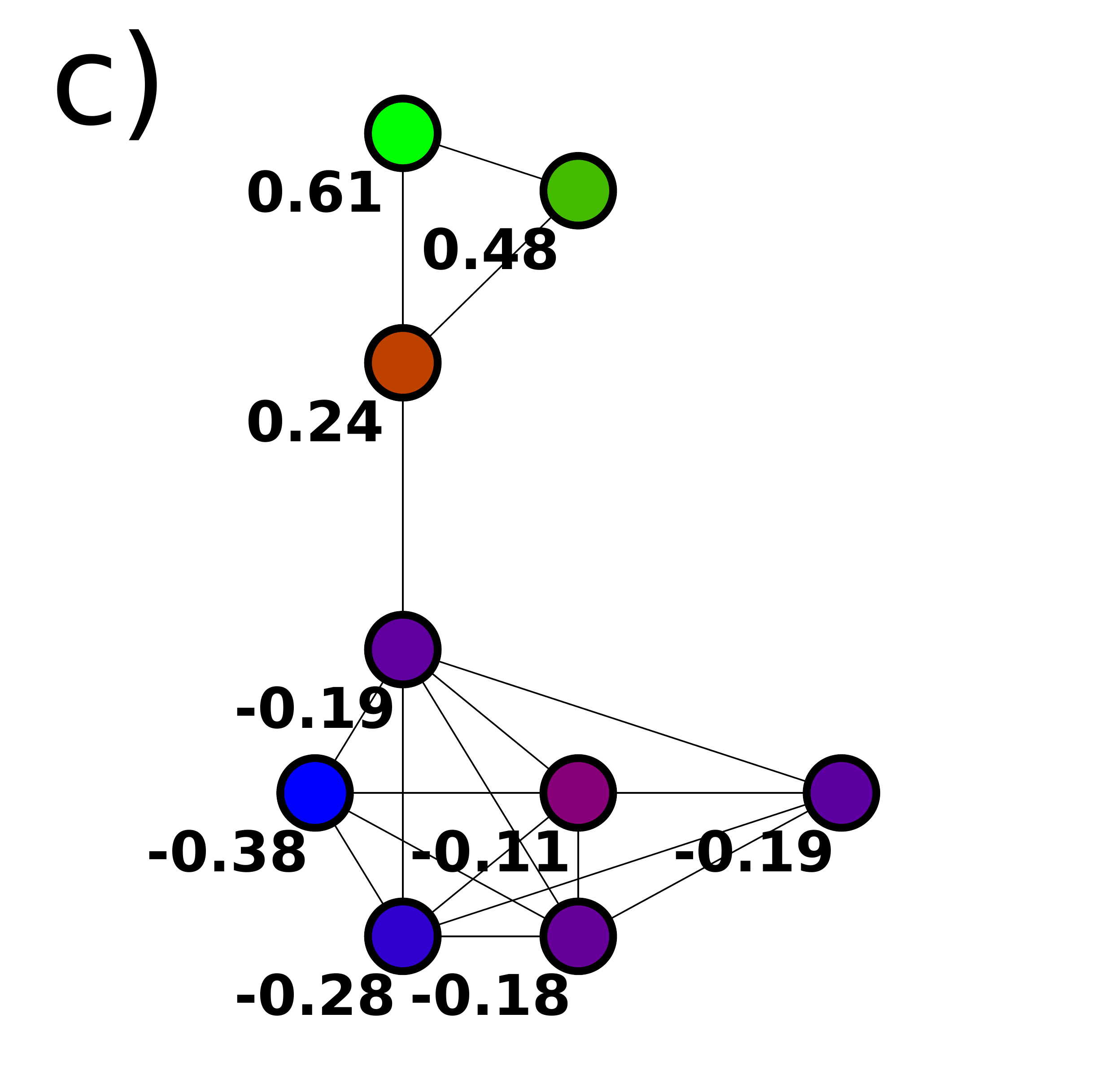}}
\vspace*{-1.5em}

\caption{Schematic illustrating the different divisions of Fiedler vector components which can occur in the RGG ensembles studied. \ref{Loc1} Shows the case when the partition is approximately even. This regime is found in solid and periodic RGGs with $d=2$. \ref{Loc2} Shows the case where most of the mass of the eigenvector is localised onto a single low degree node. In this case the value of $\mu_2$ can be strongly correlated with $\kappa_{\mathrm{min}}$. This regime is found in ER graphs as well as solid and periodic RGGs in higher dimensions. \ref{Loc3} Shows that case where the much of the mass of the eigenvector is concentrated onto a weakly connected subgraph. This case is observed in Gaussian RGGs. In both \ref{Loc2} and \ref{Loc3} we can see large fluctuations in $\mu_2$ across the ensemble as the presence of weakly-connected subgraph or low degree node is a chance event which occurs with an appreciable frequency. In cases b) and c) the values of Fiedler vector components corresponding to each node are shown. }
\label{Localisation_Compare}
\end{figure}

It is also possible to interpret the results of section \ref{dimvar} in terms of the localisation properties of the Fiedler vector. We have identified three notable cases of how the eigenvector can behave in the network ensembles studied (Figure \ref{Localisation_Compare}). Firstly, a roughly even division of the positive and negative values of the Fiedler vector components between the nodes (Figure \ref{Loc1}). Secondly, concentration of most of the eigenvectors mass onto a single node (Figure \ref{Loc2}). Finally, splitting of positive and negative components either side of a bottleneck (Figure \ref{Loc3}). The first of these occurs in low dimensional homogeneous RGGs while the latter two occur in high dimensional homogeneous RGGs where the mass of the Fiedler vector can become concentrated on the node with the lowest degree and in Gaussian RGGs where much of the mass can become concentrated on a weakly connected subgraph. 

It is not particularly surprising that $\mu_2$ shows strong correlations with $F_P$  since both $\mu_2$ and the Fiedler vector, $\underline{u}_2$ are the result of the same computation. However, as noted above, the differing behaviour of $\underline{u}_2$ for different graphs ensembles provides an interpretation of the statistical behaviour of $\mu_2$. The presence of localisation of the eigenvector demonstrates that the value of $\mu_2$ for a particular network can be influenced strongly by the edges incident to some small subset of nodes in the graph. This phenomena is discussed in \cite{atay_network_2006} where they conclude that this implies that the bulk statistical properties of a network are unlikely to be predictive of $\mu_2$. We have shown that this appears to be the case for higher dimensional RGGs with uniform node distributions where the presence of a single node with a lower degree can dramatically influence the value of $\mu_2$ (Figure \ref{Mu2_Hists}) and that this can also occur in Gaussian RGGs (Figure \ref{Gaussian_Results}) where the presence of a small weakly connected subgraph can lower the value of $\mu_2$ significantly. 

The fraction of nodes assigned to the smaller partition, $F_P$, can vary quite dramatically for different RGGs within the same ensemble (Figures  \ref{low_mu2_gauss} and \ref{high_mu2_gauss}). This indicates that the spectral clustering algorithm described in this paper may not produce meaningful results when used on different instances of RGGs drawn from the same ensemble. This is in accordance with \cite{von_luxburg_consistency_2008} where they show that the spectral clustering algorithm based on the Fiedler vector does not necessarily produce consistent results and that one can identify whether this will be the case based on the values of the lowest Laplacian eigenvalues (such as $\mu_2$).

\section{Discussion} \label{Discussion}

\subsection{Ensemble Variability in RGGs}

Node attributes and metadata such as geographic coordinates are often informative about the structure of complex networks. Gaining access to this data may be easier than obtaining the full graph structure, which may also be unnecessary if the aim is only to predict bulk properties of the graph. The aim of this study was to investigate the degree to which knowledge of node locations in RGGs gives an advantage over only having access to the distribution of node locations.

The question posed above can be rephrased as one about whether properties of RGGs are ``ensemble averageable''. In this study we have focused on the ensemble variability of $\mu_2$ since it is of interest when studying the dynamical properties of networks. We have quantified the sample-to-sample variability in $\mu_2$ by estimating the coefficient of variation of its distribution. This metric provides a measure of how robustly we can estimate $\mu_2$ for a random draw from the ensemble given knowledge of $P(\underline{X})$ and $R$. In some applications an order of magnitude estimate of the quantity of interest may be sufficient, however, for cases where we wish to estimate a quantity precisely, fluctuations greater than $5-10 \%$ of the size of the mean value may be considered significant.

Our results indicate that $CV(\mu_2)$ can exceed values of $0.3$ in ensembles with $10^5$ nodes for a range of choices of dimension and boundary conditions, including 2d RGGs with inhomogeneous node distributions (Figures \ref{fig_CV_as_d} and \ref{Gauss_CV}). The presence of a large values of $CV(\mu_2)$ for relatively large graphs demonstrates that RGG ensembles exist where $\mu_2$ is not well represented by the ensemble mean. This finding contrasts with those in \cite{kim_ensemble_2007} where they found that the extremal eigenvalues of the Laplacian for certain scale-free networks are typically ensemble averageable. In \cite{carlson_sample--sample_2011} they consider graph ensembles generated by applying a degree-sequence preserving rewiring algorithm to a range of real world networks. They found that large fluctuations in $\mu_2$ are generally observed in networks with low minimum degrees. This is consistent with our observation that $CV(\mu_2)$ is large in ensembles where $\kappa_{\mathrm{min}}$ is small, and that $\kappa_{\mathrm{min}}$ can play an important role in driving the fluctuations in $\mu_2$. 

The presence of \emph{multimodality} in the $\mu_2$ distributions of RGGs (Figures \ref{hist_d5} and \ref{hist_d10}) demonstrates that the behaviour observed in \cite{carlson_sample--sample_2011} was not just an artefact of the choices of rewiring algorithm and network data used to generate graph ensembles studied and also occurs in well studied random graph models. This multimodality was found to arise due to the strong correlation between $\mu_2$ and $\kappa_{\mathrm{min}}$. In \cite{carlson_sample--sample_2011} they hypothesize that the main factors likely to drive large fluctuations in $\mu_2$ across the ensemble are the presence of low degree nodes, the presence of `bottleneck'-like subgraphs and communities of densely connected nodes which are sparsely connected to the rest of the network. Our analysis in section \ref{Fied_Sect} indicates that this is likely to be the case for RGGs. 

One could interpret the results presented as evidence that $\mu_2$ is not necessarily the best metric to rely on when attempting to describe the bulk dynamical properties of the network; especially because it can be sensitive to microscopic properties of the network (ie. $\kappa_{min}$) in certain ensembles. Recent work in \cite{estrada_quasirandom_2017} compared RGGs generated from uniform random distributions with those where the positions are generated from `quasi-random' sequences. They found that the assortativity, spectral gap and $\mu_2$ were most strongly effected by the change in the nodal distribution. This observation was explained by noting that `quasi-random' point distributions generally appear more uniform leading to fewer clusters and holes in the RGG. This supports our finding that the presence of non-uniformity (modelled using a Gaussian node distribution) can lead to significant fluctuations in the value of $\mu_2$. 

\subsection{Implications for Node Location Knowledge in Geometric Networks}

Based on the argument made in \ref{Backgroun} (in particular see Figure \ref{RGG_Example}b) we expect knowledge of node locations in the embedding space to be of utility when $CV(\mu_2)$ is large. Precise knowledge of node locations may be possible in engineered systems such as wireless communications networks, whereas, in many other systems, such as large scale social networks, data may be noisy or inaccurate. We have identified two predominant cases where $CV(\mu_2)$ can be large: 
\begin{enumerate}[leftmargin=*]
\item{In higher dimensional RGGs with uniform node distributions $\mu_2$ can be determined by the presence of a single weakly connected node. In this case we may require detailed knowledge of both $X_i$'s for all nodes and $R$ to obtain a precise estimate of $\mu_2$.}
\item{For Gaussian RGGs we have evidence that the presence of mesoscopic subgraphs drive changes in $\mu_2$ (Figures \ref{low_mu2_gauss} and \ref{high_mu2_gauss}). This suggest that $\mu_2$ now depends on mesoscopic variation in the graph structure rather than microscopic fluctuations. This implies that knowledge of node locations may yield some predictive power, even in the case where this knowledge is not precise. In particular, knowledge of node locations and connectivity in low density regions is likely to be of most utility. 

A similar conclusion applies for RGGs in high dimensions with solid boundaries where fluctuations in $\mu_2$ are driven by chance events occuring at the boundary of the domain. This suggests that, when it comes to predicting $\mu_2$ and related metrics that \emph{knowledge of node locations at the boundary will be of most utility} }
\end{enumerate}

For uniform node distributions in low dimensions the value of $CV(\mu_2)$ is relatively small (Figures \ref{CV_as_d_Periodic} and \ref{CV_as_d_Solid}). This suggests that knowledge of network properties such as $\mu_2$ for a particular graph might be obtained with a reasonable precision simply by sampling from an ensemble with an equivalent node distribution and connection radius. In this case, additional knowledge of $\underline{X}$ over $P(\underline{X})$ may not be of significant utility. In \cite{butts_predictability_2003} they use graph entropy in order to quantify the predictive advantage gained from knowledge of $\underline{X}$ in 2d Soft RGGs. They find that knowledge of node locations can account for the vast majority of network structure. In this work we have shown that, even when knowledge of $\underline{X}$ gives us full knowledge of the graph structure (as in the RGGs), structural and dynamically relevant properties may vary little between different graphs drawn from the same ensemble. In contrast to this, large scale social networks can typically be represented with a relatively low number of dimensions \cite{bonato_dimensionality_2014} with individuals often being distributed inhomogeneously in the important dimensions such as those of geographic space \cite{butts_geographical_2012}. The observations above suggest that some care is required when linking dynamical properties of an ensembles these systems (e.g. having information about population densities rather than individual microdata) to those of of a particular instance.

\subsection{Future Prospects}

In this work we have characterised ensemble variability in RGGs in terms of the variability in $\mu_2$. A complementary approach to quantifying the degree of variability or `topological uncertainty' in a random graph ensemble is through the notion of graph entropy. This has recently been studied in RGGs \cite{coon_topological_2016,coon2018entropy,badiu2018distribution}. An interesting avenue of future research will be to explore the degree to which topological uncertainty in the graph structure (as measured by entropy or other metrics) coincides with large variability in the dynamical properties. Furthermore, in Soft RGGs we can also consider entropies which are conditioned on the internode distances \cite{coon2018entropy} or node locations \cite{halu_emergence_2014}. Studying these along with the level of ensemble variability in graph properties would allow us to understand how informative node locations are in graphs with probabilistic connection functions.

The adjacency and Laplacian matrices of RGGs and Soft RGGs take the form of Euclidean Random Matrices (ERMs) in which matrix entries are functions of the random positions. Results exist for the expected values of their eigenvalues in the asymptotic limit \cite{bordenave_eigenvalues_2008} and it has also been shown that the spectra of these matrices have certain properties in common with well studied ensembles from Random Matrix Theory \cite{dettmann_spectral_2017}. However, to the authors knowledge, no analytic results exist concerning the distributional properties of individual eigenvalues in these systems. We have shown that these distributions can behave in a non-trivial manner in response to changes in the system parameters (Figure \ref{Mu2_Hists}). Consequently, we believe that it is of interest to obtain a theoretical understanding of the factors leading this. For example, it would be of interest to extend theoretical techniques in order find an expression for $CV(\mu_2)$ and if possible provide a full characterization of $\mathbb{P}(\mu_2)$.

A wide range of real world systems can be modelled with ensembles of RGGs and their generalizations \cite{barthelemy_spatial_2011,barnett_spatially_2007,butts_spatial_2011,butts_interorganizational_2012,daraganova_networks_2012,odea_spreading_2013,lo_geometric_2015,hackl2017generation}. These consist of both spatially embedded systems and graphs where the nodes posses other attributes which are relevant to tie formation. Providing limits on how much we can predict about the properties of these graphs given the available data is essential for quantifying uncertainty in results and understanding level of data required for a specific application. This study takes the first step in this direction by attempting to understand the factors which drive variability in dynamically relevant properties of RGGs. Our results suggest that the amount we can predict about a graph given node locations or a distribution can vary significantly for different RGG ensembles with node-location information being particularly useful in the context of inhomogenous node distributions. This suggests that knowledge of the generative process that leads to the formation of a network or detailed knowledge of the relevant node attributes and their corresponding distributions may be required when attempting to estimate graph properties without full access to the graph structure.

\section*{Acknowledgements}

This work has been supported by EPSRC grants EP/L016613/1 (Centre for Doctoral Training in the Mathematics of Planet Earth) and EP/N014529/1 (Centre for the Mathematics of Precision Healthcare). The authors thank Till Hoffman and Spencer Wilson for their contributions to the code base and useful discussions. Thanks to Thomas Gibson and Carl Dettmann for their comments on the manuscript. Simulations were performed using Imperial College High Performance Computing cluster: Imperial College Research Computing Service,  DOI: \href{http://doi.org/10.14469/hpc/2232}{10.14469/hpc/2232}.

\bibliographystyle{apsrev4-1}
\bibliography{bibliography}

\clearpage


\section*{Supplementary Information}
\beginsupplement

\setcounter{section}{0}

\section{Simulation Methodology} \label{EV_Computation_Methods}

RGGs were generated using code in the Python programming language. Adjacency matrices for the graphs were stored in sparse matrix formats. A significant speed up in the procedure used to generate RGGs was obtained by using the Cython package for Python. In addition to this, efficient computation of nearest-neighbour distance was implemented using a KD-Tree based data structure for storing the node positions. Using this methodology it was possible to efficiently draw adjacency matrices for RGGs with sizes of up to $N \approx 10^5 $. The eigenvalues of large sparse Laplacian matrices were computed using functions from the ARPACK toolbox in the Python programming language (For details see: \href{https://docs.scipy.org/doc/scipy-0.18.1/reference/tutorial/arpack.html}{https://docs.scipy.org/doc/scipy-0.18.1/reference/tutorial/arpack.html}).

The code used for constructing RGGs and computing their properties is available at:
\\

\href{https://github.com/MGarrod1/rgg_ensemble_analysis}{https://github.com/MGarrod1/rgg\_ensemble\_analysis}
\\

The github repository contains example scripts for producing RGGs and a subset of figures in the manuscript. The example scripts allow the user to produce data from a subset of the figures which consider RGGs with smaller values of $N$. It is principle possible to reproduce $CV(\mu_2)$ values obtained for RGG ensembles with different parameters, however, simulations with $N \approx 10^5$ require a significant run time.

\section{Generation of RGGs with given expected Degrees} \label{meandeg_alg}

\textbf{Estimation of required radius via sampling from the distance distribution.} For RGGs with general node distribution and boundary conditions it is not straightforward to analytically compute the value of $R$ required for graphs with a given $\kappa$. Consequently, it is necessary to have a systematic approach that allows us to identify values of $R$ which correspond to desired choices of $\kappa$. In order to do this we use a method which relies on fixing the number of edges in the network (a similar approach is used in \cite{donges_backbone_2009}). 
The mean degree of a network is related to the number of edges, $E$, by:
\begin{equation} \label{Edge_Dense}
\kappa = \frac{2E}{N} . 
\end{equation}
Given a set of points $X_1,X_2,...,X_N$ in some domain we can compute a sorted set of pairwise Euclidean distances: $\delta_1 \leq \delta_2 \leq ... \leq \delta_P$ where $P = \frac{N(N-1)}{2}$. To generate an RGG with $E$ edges we require that at least $E$ pairs of nodes lie within a Euclidean distance $R$ of each other. Consequently, the connection radius required to generate a graph containing $E$ edges will satisfy the inequality:
\begin{equation}
\delta_E < R < \delta_{E+1} . 
\end{equation}
Choosing a value of $R$ which falls within this range will allow us to construct an RGG with a mean degree of $\kappa$ for the specified set of positions. 

Choosing the connection radius based on the distance distribution allows us to generate an RGG with the desired mean degree without having to explicitly generate the graph first. This approach also has the advantage that it allows us to identify the desired value of $R$ for any choice of domain or node distribution for which we can compute pairwise distances. 

\textbf{Subsampling approach for larger networks.} For a network containing $N$ nodes we must store an array of $\frac{N(N-1)}{2}$ pairwise distances. This will become computationally infeasible for large $N$. Consequently, for a given $N$ and $\kappa$ we estimate $R$ by drawing a smaller sample from the node distribution of $M=1000$ positions which gives us access to an array of $499,500$ samples from the distribution of distances. Given this we can approximate the connection radius with:
\begin{equation}
R \approx \delta_{ \lfloor E' \rfloor } ,
\end{equation}
where $\lfloor x \rfloor $ denotes the floor function and $E' = \frac{2E}{N(N-1)} \frac{M(M-1)}{2} $ and we have ordered the distances according to their size. We have used numerical simulations to confirm that the above methodology produces RGG ensembles with mean degrees close to the desired mean degree. 

In order to robustly estimate the required value of $R$ we perform multiple simulations using the methodology described above and take the mean value. Unless otherwise specified the values of $\kappa$ reported in the manuscript correspond to the desired $\kappa$ used as an input for the algorithm as opposed to the empirically observed mean degree for the graph ensemble. For sufficiently large graphs and large sample sizes these values should be close to each other. 

\section{Analytic Formula for the Algebraic Connectivity} \label{Analytic_Derivation}

In this section we present an analytic formula to approximate the average algebraic connectivity for RGGs with periodic boundaries in $[0,1]^d$. The formula derived applies for RGGs with uniform node distributions, however, the techniques used could also be applied to both the non-uniform node distributions and networks with soft connection functions. 

\subsection{Derivation}

In order to derive an analytic expression for $\mu_2$ we consider the more general case of Soft Random Geometric Graphs \cite{dettmann_random_2016}. For these graphs connections between nodes are now made probabilistically given the node positions. A connection between nodes at positions $\underline{x}_i$ and $\underline{x}_j$ is made with probability $\gamma(|\underline{r}|)$, where $\underline{r} = \underline{x}_i - \underline{x}_j$. We will refer to this function as the connection function. 

The adjacency matrix of a Soft RGG takes the form of a Euclidean Random Matrix (ERM). An ERM, $M$, is a matrix where the entries are functions of positions in Euclidean space. This means that the entries take the form:
\begin{equation}
M_{ij} = F( | \underline{x}_i - \underline{x}_j| ) , 
\end{equation}
where $F$ is some measurable mapping and $\underline{x}_i$ are positions in some domain. In \cite{bordenave_eigenvalues_2008} it is shown that the eigenvalues of an ERM can be expressed in terms of the Fourier coefficients of the connection function. In particular, the eigenvalues of the adjacency matrix of a spatial network with a uniform distribution of points on the unit torus take the form:
\begin{equation}
\lambda_i = N \hat{\gamma}( \underline{k} ) ,
\end{equation}
where $\hat{\gamma}( \underline{k} )$ are the Fourier coefficients of the connection function. These can be expressed as:
\begin{equation} \label{Fourier1}
\hat{\gamma}( \underline{k} ) = \int_{ \mathbb{R}^n } \gamma(\underline{r}) e^{- 2 \pi i \underline{k} \cdot \underline{x} } d \underline{r} ,
\end{equation}
where $\underline{k} \in \mathbb{Z}^d $. We note here that for a non-uniform distribution of positions the integration with respect to $d \underline{r}$ can be replaced by an integration with respect to the probability measure of interest. 

In \cite{sardellitti_optimal_2012} these results are applied to derive an expression for the  algebraic connectivity of an RGG with points in a two dimensional periodic domain. We now generalize this result to the case of an RGG in a $d$ dimensional periodic domain. 

In order to determine the second smallest eigenvalue of $L$ we note that the eigenvalues of the matrix with entries $\delta_{ij}K_j$ are simply the degrees of the network. Taking the same approach as \cite{sardellitti_optimal_2012} we note that in the asymptotic limit these will tend to the mean degree of the network. 
Consequently, for large $N$, the eigenvalues of $L$ can be approximated by:
\begin{equation} \label{TwoEigs}
\mu_i \approx \kappa - \lambda_i , 
\end{equation}
where $\lambda_i$ are the eigenvalues of the adjacency matrix. From this we can observe that the second smallest eigenvalues of $L$ can be determined from the second largest eigenvalue of $A$.

In \cite{sardellitti_optimal_2012} they show that the magnitude of the Fourier coefficients is a decreasing function of the magnitude of the wavevector $|k|$ for $d=2$. We will assume that this result extends to the case of higher dimensions. Given this, we can obtain the algebraic connectivity by computing the Fourier coefficient corresponding to $\underline{k} = (0,...,0,1)$ (or without loss of generality, any other vector for which $|k|=1$). Furthermore, setting $\underline{k} = \underline{0}$ gives:
\begin{equation}
\hat{\gamma}( \underline{0} ) = N \int_{ \mathbb{R}^n } \gamma(\underline{r}) d \underline{r} = \kappa .
\end{equation}
From this we see that $\mu_1 = \kappa - \kappa = 0$ which is the expected result for the Laplacian matrix. 

We now consider the case of an RGG with uniform node distribution in the domain $[0,1)^d $ with periodic boundary conditions. The periodic boundary conditions are used in order to remove boundary effects which simplifies the calculation.  For an RGG with a connection radius $R$ the connection function takes the form:
\begin{equation}
\gamma(\underline{r} ) = \begin{cases}
1 \quad \mathrm{if} \quad |\underline{r}| \leq R ,\\
0 \quad \mathrm{otherwise}
\end{cases} . 
\end{equation}
Substituting this into \eqref{Fourier1}, we see that for this connection function the eigenvalues become equivalent to the Fourier coefficients of the indicator function of the ball of radius $R$ in $d$ dimensions. We note that this of course assumes that $R$ is smaller than the domain size. The eigenvalues of the adjacency matrix therefore take the form:
\begin{equation}
\gamma(\underline{k}) = \int_{ |\underline{r}| \leq R} e^{- 2 \pi i \underline{k} \cdot \underline{r} } d \underline{r} . 
\end{equation}
This integral is difficult to perform for a general $\underline{k}$, however, for our purposes it is sufficient to perform the integration along the $d th$ axis of the unit ball. Setting $\underline{k} = \underline{k}_{d,k} = (0,...,0,k)$ allows us to obtain:
\begin{equation}
\hat{\gamma}(\underline{k}_{d,k}) = \int_{ |\underline{x}| \leq R} e^{- 2 \pi i k r_d } d r_1 ... d r_d . 
\end{equation}
Since the integrand now only depends on $r_d$ we can write this integral in the form:
\begin{equation}
\hat{\gamma}(\underline{k}_{d,k}) = \int_{-R}^{R} e^{- 2 \pi i k r_d } \int_{r_1^2 + ... + r_{d-1}^2 \leq R^2 - r_d^2} d r_1 ... d r_{d-1} d r_{d} . 
\end{equation}
The second integral corresponds to the volume of the sphere of radius $\sqrt[]{R^2 -r_d^2}$ in $d-1$ dimensions. The volume of a $d$ dimensional sphere of radius $\rho$ takes the form: 
\begin{equation}
V_d(\rho) = \frac{\pi^{\frac{d}{2}}}{\Gamma( \frac{d}{2} + 1 )} \rho^d . 
\end{equation}
Therefore we obtain:
\begin{equation}
\hat{\gamma}(\underline{k}_{d,k}) = \int_{-R}^{R} e^{- 2 \pi i k r_d } \frac{\pi^{\frac{d-1}{2}}}{\Gamma( \frac{d-1}{2} + 1 )} (R^2 - r_d^2)^{\frac{d-1}{2}} d r_d . 
\end{equation}
We now make the change of variables $r_d = R cos(\theta) $ and set $k=1$. After some re-arrangement we obtain:
\begin{equation} \label{FourierDef}
\hat{\gamma}(\underline{k}_{d,1}) =   \frac{\pi^{\frac{d-1}{2}}}{\Gamma( \frac{d-1}{2} + 1 )} \int_{0}^{\pi} R^{d} \sin^{d} (\theta) e^{- 2 \pi i R cos(\theta) }  d \theta .
\end{equation}
In order to solve this integral we make use of the following integral representation of the Bessel function of the first kind (see expression 9.1.20 on pg. 360 of \cite{abramowitz1964handbook}): 
\begin{equation}
J_{p}(x) = \frac{\big( \frac{x}{2} \big)^p}{\sqrt[]{\pi} \Gamma\big( p + \frac{1}{2} \big) } \int_0^\pi \sin^{2p}(\theta) \cos( x cos(\theta) ) d \theta .
\end{equation}
The integral above can be split into two integrals by splitting the complex exponential into separate $\sin$ and $\cos$ terms. Noting that the second integral goes to zero due to symmetry, we obtain:
\begin{equation} \label{BeselDef}
J_{p}(x) = \frac{\big( \frac{x}{2} \big)^p}{\sqrt[]{\pi} \Gamma\big( p + \frac{1}{2} \big) } \int_0^\pi \sin^{2p}(\theta) e^{-i x cos(\theta) } d \theta .
\end{equation}
By comparison of \eqref{BeselDef} and \eqref{FourierDef} we find that:
\begin{equation}
\hat{\gamma}(\underline{k}_{d,1}) = R^{\frac{d}{2}} J_{\frac{d}{2}} (2 \pi R ) . 
\end{equation}
Substitution into \eqref{TwoEigs} allows us to obtain:
\begin{equation} \label{Theory}
\mu_2 \approx \kappa - N R^{\frac{d}{2}} J_{\frac{d}{2}} (2 \pi R )
\end{equation} 

The techniques used to derive \ref{Theory} can in principle be extended to the case of general domains and connection functions. This can be achieved by choosing the desired connection function in equation \eqref{Fourier1} and then integrating w.r.t the desired node distribution in the corresponding domain. In many cases the resulting integral may not be tractable, however, this equation still provides some insight into the factors which effect the average value of $\mu_2$. Furthermore, it may be possible to extend the ideas in \cite{bordenave_eigenvalues_2008} in order to compute different eigenvalues from the spectrum of $L$.

\section{Standard Error in the Coefficient of Variation} \label{CV_SE}

In order to estimate the statistical uncertainty in the coefficient of variation (CV) we require an estimate for the standard error in the variance of a sample. For a sample $\underline{y} = (y_1,y_2,...y_N)$ with mean $\langle y \rangle $, the standard error on the variance of the sample is given by: \cite{wonnapinij_previous_2010}
\begin{equation} \label{CV_ER}
\sigma_{SE}(Var) = \sqrt[]{\frac{1}{N} \bigg(  D_4 - \frac{N-3}{N-1} \sigma^4  \bigg)} , 
\end{equation}
where $D_4$ is the fourth central moment of the sample. An unbiased estimator for this quantity can be calculated from:
\begin{equation}
D_4 = \frac{(N-1)}{N^3} \bigg( (N^2 -3N + 3) M_4 + 3(2N - 3 ) M_2^2 \bigg) , 
\end{equation}
where:
\begin{equation}
M_j = \frac{1}{N} \sum_{i=1}^N \big( y_i - \langle y \rangle )^j \big) .
\end{equation}
Combining \eqref{CV_ER} with the standard error in the mean using traditional error propagation techniques allows us to obtain an estimate for the uncertainty on the CV. The results obtained using \eqref{CV_ER} were validated by comparison with an estimate of the standard error computed via a standard bootstrapping method.

\section{Effect of Full Connectivity on Fluctuation in $\mu_2$.} \label{Percol_Sect}

\begin{figure}[h!]
\centering
\subfloat[]{\label{supp_1}\includegraphics[width=.95\textwidth]{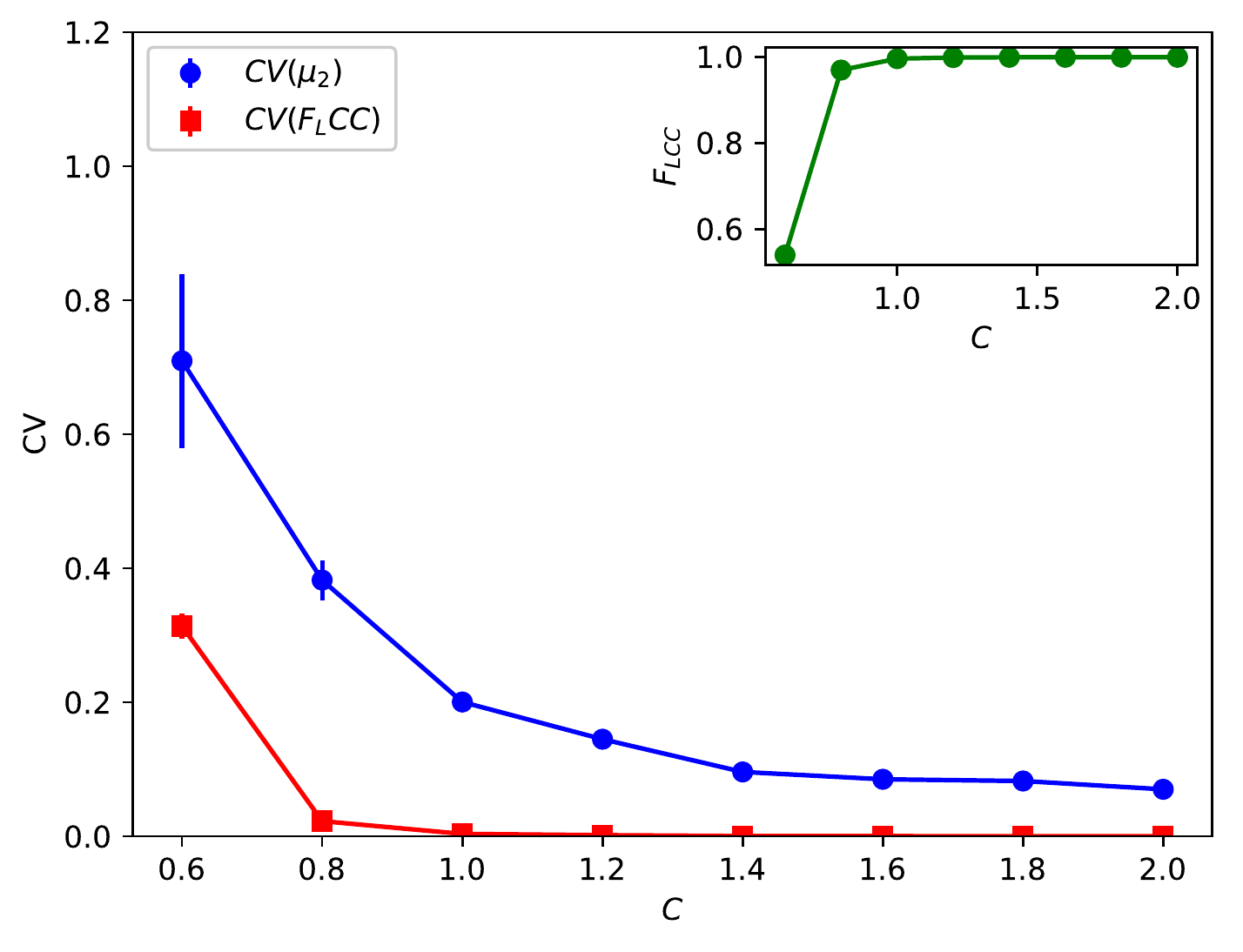}}

\caption{Decreasing towards and going below the connectivity threshold causes large fluctuations in $\mu_2$. Plot showing the behaviour of $CV(\mu_2)$ and $CV(N_{LCC})$ as a function of $C$ for periodic RGG ensembles with $d=2$, $N=1000$. Each data point corresponds to 100 samples from the ensemble. The inset shows the average value of $F_{LCC}$ for the same range of parameters. }
\label{percol_effects}
\end{figure}

In the main text we focus on RGG ensembles for which graphs are connected or close to being so ($N_{LCC} \approx N$). For smaller values of $C$ (or $\kappa$) we obtain graph ensembles in which the size of the LCC will fluctuate. This introduces an additional source of randomness into the system. Figure \ref{percol_effects} shows how $CV(\mu_2)$ and $CV(N_{LCC})$ behave as we change reduce $C$ below the threshold of full connectivity at $C=1$. $CV(\mu_2)$ becomes large for systems below the connectivity threshold, however, it begins to increase well before reaching $C=1$ suggesting that sparser and more fragmented networks are naturally more variable in their dynamical properties.

\section{Effect of Minimum Degree Fluctuations in Gaussian RGGs} \label{Gauss_Min_Fluct}

\begin{figure}[]
\centering
\subfloat[]{\label{Gauss_CV}\includegraphics[width=.95\textwidth]{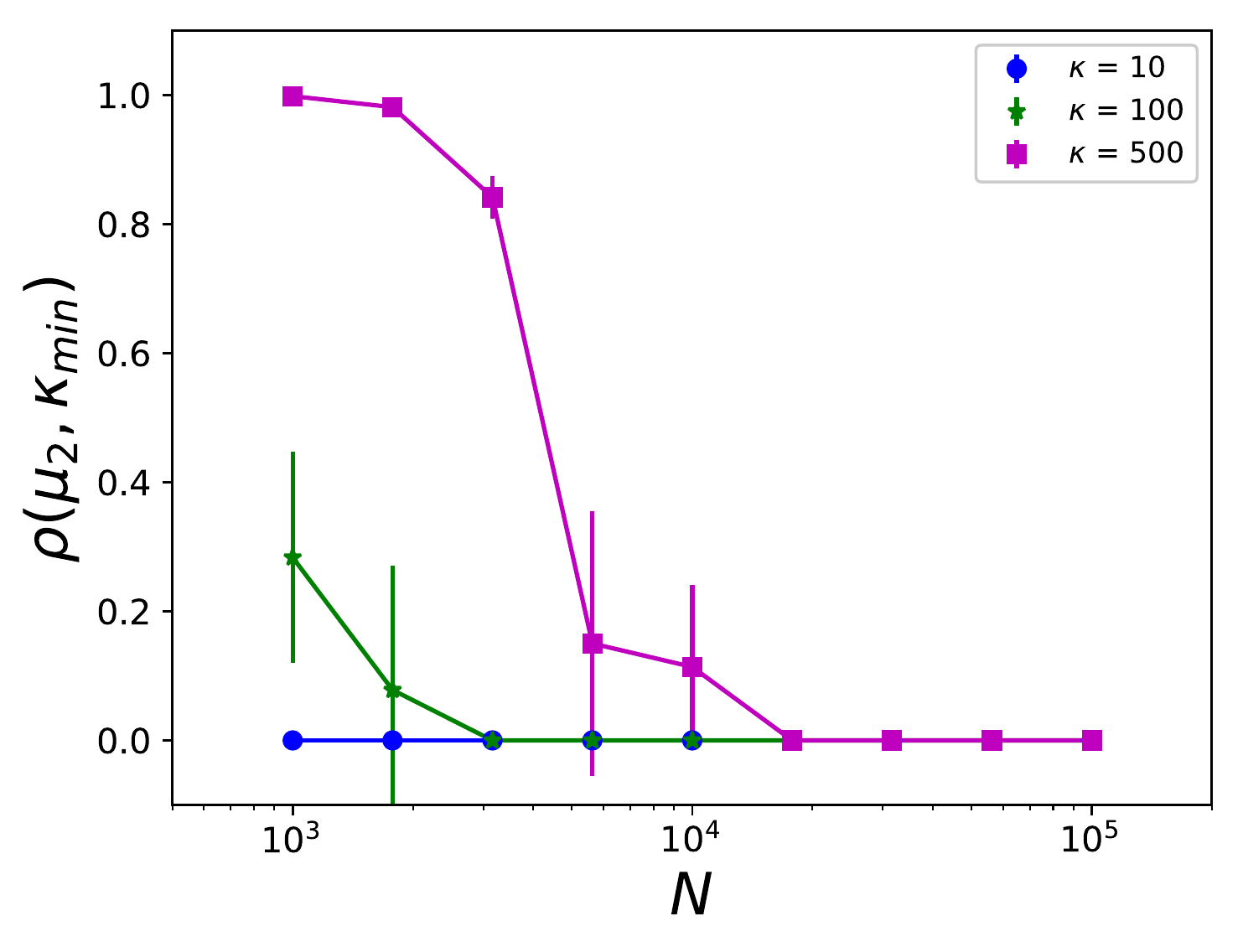}}

\caption{$\kappa_{\mathrm{min}}$ can also drive variability in $\mu_2$ in Gaussian RGGs. Plot showing $\rho(\mu_2,\kappa_{\mathrm{min}})$ as a function of $N$ for Gaussian RGGs with different values of the $\kappa$. For systems with larger $N$ the value of $\rho(\mu_2,\kappa_{\mathrm{min}})$ drops to zero as all RGGs in these ensembles will have $\kappa_{\mathrm{min}}=1$ with high probability. The value of $\rho(\mu_2,\kappa_{\mathrm{min}})$ becomes significant in smaller Gaussian RGGs with larger values of $\kappa$ where it is possible to sample RGGs with a range of $\kappa_{\mathrm{min}}$ values. Each data point corresponds to 100 samples from the corresponding ensembles.}
\label{Gaussian_kmin_corr}
\end{figure}

In the main text we illustrate that fluctuations in $\mu_2$ for Gaussian RGG ensembles are typically driven by the presence of weakly connected subgraphs in low node density regions of the domain. This is the case in the regime with large $N$ and small $\kappa$, however, it is also possible to observe cases in Gaussian RGGs where fluctuations in the $\kappa_{\mathrm{min}}$ drive those in $\mu_2$ (Figure \ref{Gaussian_kmin_corr}). This demonstrates that large values of $\rho(\mu_2,\kappa_{\mathrm{min}})$ in a particular graph ensemble are not necessarily driven by the dimensionality of the system (as is the case in homogeneous systems).

\section{Tables of Parameters} \label{Tables}

Tables of connection radii values generated using the approach described in section \ref{meandeg_alg} for the different figures in the manuscript are shown below. Where relevant we also report the value of $C=\frac{\kappa}{\log(N)}$ where $\kappa$ is the expected degree for homogeneous RGGs. For ensembles in $[0,1]^d$ (Tables \ref{fig2_tab},\ref{fig3_tab} and \ref{fig4_tab}) we use P to denote ensembles with periodic boundaries and S to denote ensembles with solid boundaries. Table \ref{fig5_tab} corresponds to input parameters for Gaussian RGGs for which no boundary conditions were applied.

\begin{table}[hbt!]
\begin{tabular}{|l|l|l|l|}
\hline
Boundary & $R$            & d  & N    \\ \hline
P        & 0.1261831574 & 2  & 1000 \\ \hline
S        & 0.1337362242 & 2  & 1000 \\ \hline
P        & 0.2280560543 & 3  & 1000 \\ \hline
S        & 0.2516758615 & 3  & 1000 \\ \hline
P        & 0.3174380722 & 4  & 1000 \\ \hline
S        & 0.3628587046 & 4  & 1000 \\ \hline
P        & 0.3942367888 & 5  & 1000 \\ \hline
S        & 0.4611800289 & 5  & 1000 \\ \hline
P        & 0.4615454243 & 6  & 1000 \\ \hline
S        & 0.5525702376 & 6  & 1000 \\ \hline
P        & 0.5219885978 & 7  & 1000 \\ \hline
S        & 0.6359722925 & 7  & 1000 \\ \hline
P        & 0.5779606321 & 8  & 1000 \\ \hline
S        & 0.7170924116 & 8  & 1000 \\ \hline
P        & 0.6295518242 & 9  & 1000 \\ \hline
S        & 0.7898305619 & 9  & 1000 \\ \hline
P        & 0.6777076661 & 10 & 1000 \\ \hline
S        & 0.8583408529 & 10 & 1000 \\ \hline
P        & 0.7237600487 & 11 & 1000 \\ \hline
S        & 0.9254288384 & 11 & 1000 \\ \hline
P        & 0.7673975833 & 12 & 1000 \\ \hline
S        & 0.9862521976 & 12 & 1000 \\ \hline
P        & 0.8091803122 & 13 & 1000 \\ \hline
S        & 1.0442485441 & 13 & 1000 \\ \hline
P        & 0.8495710964 & 14 & 1000 \\ \hline
S        & 1.1010651214 & 14 & 1000 \\ \hline
\end{tabular}
\caption{Table of Parameters used to produce data in figure 2. Connection radii were estimated by taking the mean of 5 samples from the algorithm presented in \ref{meandeg_alg}.}
 \label{fig2_tab}
\end{table}

\begin{table}[hbt!] 
\begin{tabular}{|l|l|l|l|l|}
\hline
d  & N     & $C$ & $R$           & Boundary \\ \hline
2  & 10000 & 1.5                        & 0.0211050444 & P        \\ \hline
5  & 10000 & 1.5                        & 0.1922841591 & P        \\ \hline
10 & 10000 & 1.5                        & 0.4719028713 & P        \\ \hline
\end{tabular}
\caption{Table of Parameters used to produce the data for figure 3. Connection radii were estimated by taking the mean of 10 samples from the algorithm presented in \ref{meandeg_alg}.}
\label{fig3_tab}
\end{table}

\begin{table}[hbt!] 
\begin{tabular}{|l|l|l|l|l|}
\hline
d  & N      & $C$ & $R$            & Boundary \\ \hline
2  & 100000 & 2                          & 0.0084089067 & S        \\ \hline
3  & 100000 & 2                          & 0.0389121361 & S        \\ \hline
4  & 100000 & 2                          & 0.0848507036 & S        \\ \hline
5  & 100000 & 2                          & 0.1401172365 & S        \\ \hline
6  & 100000 & 2                          & 0.2004778726 & S        \\ \hline
7  & 100000 & 2                          & 0.261449855  & S        \\ \hline
8  & 100000 & 2                          & 0.32156097   & S        \\ \hline
9  & 100000 & 2                          & 0.38141171   & S        \\ \hline
10 & 100000 & 2                          & 0.4410849621 & S        \\ \hline
2  & 100000 & 7                          & 0.0160486294 & S        \\ \hline
3  & 100000 & 7                          & 0.0588770826 & S        \\ \hline
4  & 100000 & 7                          & 0.1180306656 & S        \\ \hline
5  & 100000 & 7                          & 0.1837423324 & S        \\ \hline
6  & 100000 & 7                          & 0.2508714077 & S        \\ \hline
7  & 100000 & 7                          & 0.3164284687 & S        \\ \hline
8  & 100000 & 7                          & 0.3827082921 & S        \\ \hline
9  & 100000 & 7                          & 0.4462617069 & S        \\ \hline
10 & 100000 & 7                          & 0.5097059479 & S        \\ \hline
2  & 100000 & 15                         & 0.0237815453 & S        \\ \hline
3  & 100000 & 15                         & 0.0767300043 & S        \\ \hline
4  & 100000 & 15                         & 0.1437831689 & S        \\ \hline
5  & 100000 & 15                         & 0.2160864297 & S        \\ \hline
6  & 100000 & 15                         & 0.2875199113 & S        \\ \hline
7  & 100000 & 15                         & 0.3593604744 & S        \\ \hline
8  & 100000 & 15                         & 0.4264113804 & S        \\ \hline
9  & 100000 & 15                         & 0.4952014366 & S        \\ \hline
10 & 100000 & 15                         & 0.5583207176 & S        \\ \hline
2  & 100000 & 1.5                        & 0.0073697699 & P        \\ \hline
3  & 100000 & 1.5                        & 0.0342700894 & P        \\ \hline
4  & 100000 & 1.5                        & 0.0764912423 & P        \\ \hline
5  & 100000 & 1.5                        & 0.1257266863 & P        \\ \hline
6  & 100000 & 1.5                        & 0.1782134719 & P        \\ \hline
7  & 100000 & 1.5                        & 0.232484686  & P        \\ \hline
8  & 100000 & 1.5                        & 0.2853110262 & P        \\ \hline
9  & 100000 & 1.5                        & 0.3332698942 & P        \\ \hline
10 & 100000 & 1.5                        & 0.3816671501 & P        \\ \hline
2  & 100000 & 2                          & 0.0085886599 & P        \\ \hline
3  & 100000 & 2                          & 0.0382409096 & P        \\ \hline
4  & 100000 & 2                          & 0.0826780317 & P        \\ \hline
5  & 100000 & 2                          & 0.1351192459 & P        \\ \hline
6  & 100000 & 2                          & 0.1872363186 & P        \\ \hline
7  & 100000 & 2                          & 0.2427156333 & P        \\ \hline
8  & 100000 & 2                          & 0.2943532808 & P        \\ \hline
9  & 100000 & 2                          & 0.3474842206 & P        \\ \hline
10 & 100000 & 2                          & 0.394577478  & P        \\ \hline
2  & 100000 & 2.5                        & 0.009566736  & P        \\ \hline
3  & 100000 & 2.5                        & 0.0416235599 & P        \\ \hline
4  & 100000 & 2.5                        & 0.0869156497 & P        \\ \hline
5  & 100000 & 2.5                        & 0.1396236529 & P        \\ \hline
6  & 100000 & 2.5                        & 0.1952175328 & P        \\ \hline
7  & 100000 & 2.5                        & 0.2512468638 & P        \\ \hline
8  & 100000 & 2.5                        & 0.3035246822 & P        \\ \hline
9  & 100000 & 2.5                        & 0.3529659844 & P        \\ \hline
10 & 100000 & 2.5                        & 0.4032246977 & P        \\ \hline
\end{tabular}
\caption{Table of Parameters used to produce the data for figure 4. Connection radii were estimated by taking the mean of 10 samples from the algorithm presented in \ref{meandeg_alg}.}
\label{fig4_tab}
\end{table}

\begin{table}[hbt!]
\begin{tabular}{|l|l|l|l|}
\hline
d & N      & $\kappa$ & r            \\ \hline
2 & 1000   & 10                         & 0.2018258431 \\ \hline
2 & 1000   & 100                        & 0.6495272068 \\ \hline
2 & 1000   & 500                        & 1.6682958318 \\ \hline
2 & 1778   & 10                         & 0.1481204375 \\ \hline
2 & 1778   & 100                        & 0.4823639762 \\ \hline
2 & 1778   & 500                        & 1.1549497454 \\ \hline
2 & 3162   & 10                         & 0.1127501063 \\ \hline
2 & 3162   & 100                        & 0.3601397921 \\ \hline
2 & 3162   & 500                        & 0.8390884587 \\ \hline
2 & 5623   & 10                         & 0.0858341245 \\ \hline
2 & 5623   & 100                        & 0.2670867356 \\ \hline
2 & 5623   & 500                        & 0.6025613913 \\ \hline
2 & 10000  & 10                         & 0.063283408  \\ \hline
2 & 10000  & 100                        & 0.1987662124 \\ \hline
2 & 10000  & 500                        & 0.4551119017 \\ \hline
2 & 17782  & 10                         & 0.0474059169 \\ \hline
2 & 17782  & 100                        & 0.1518550767 \\ \hline
2 & 17782  & 500                        & 0.345185169  \\ \hline
2 & 31622  & 10                         & 0.0356330017 \\ \hline
2 & 31622  & 100                        & 0.1127553845 \\ \hline
2 & 31622  & 500                        & 0.2508205296 \\ \hline
2 & 56234  & 10                         & 0.0268741945 \\ \hline
2 & 56234  & 100                        & 0.084723442  \\ \hline
2 & 56234  & 500                        & 0.189669697  \\ \hline
2 & 100000 & 10                         & 0.0202019497 \\ \hline
2 & 100000 & 100                        & 0.0639243444 \\ \hline
2 & 100000 & 500                        & 0.1426389485 \\ \hline
\end{tabular}
\caption{Table of parameters used to produce the data for figure 5 in the manuscript. Connection radii were estimated by taking the mean of 10 samples from the algorithm presented in \ref{meandeg_alg}.}
\label{fig5_tab}
\end{table}

\clearpage

\end{document}